\documentclass[trackchanges, twocolumn]{aastex701}

\begin{document}

\title{
ATLAS. IV.\\
A JWST+MUSE Demographic Study of Ly$\alpha$ Profiles in Little Red Dots
}

\author[orcid=0009-0004-0381-7216]{Yuta Kageura}
\affiliation{Institute for Cosmic Ray Research, The University of Tokyo, 5-1-5 Kashiwanoha, Kashiwa, Chiba 277-8582, Japan}
\affiliation{Department of Physics, Graduate School of Science, The University of Tokyo, 7-3-1 Hongo, Bunkyo, Tokyo 113-0033, Japan}
\email[show]{kageura@icrr.u-tokyo.ac.jp}
\correspondingauthor{Yuta Kageura}

\author[orcid=0000-0002-1049-6658]{Masami Ouchi}
\affiliation{National Astronomical Observatory of Japan, 2-21-1 Osawa, Mitaka, Tokyo 181-8588, Japan}
\affiliation{Institute for Cosmic Ray Research, The University of Tokyo, 5-1-5 Kashiwanoha, Kashiwa, Chiba 277-8582, Japan}
\affiliation{Astronomical Science Program, Graduate Institute for Advanced Studies, SOKENDAI, 2-21-1 Osawa, Mitaka, Tokyo 181-8588, Japan}
\affiliation{Kavli Institute for the Physics and Mathematics of the Universe (WPI), The University of Tokyo, 5-1-5 Kashiwanoha, Kashiwa, Chiba 277-8583, Japan}
\email{ouchims@icrr.u-tokyo.ac.jp}

\author[orcid=0009-0006-6763-4245]{Hiroto Yanagisawa}
\affiliation{Institute for Cosmic Ray Research, The University of Tokyo, 5-1-5 Kashiwanoha, Kashiwa, Chiba 277-8582, Japan}
\affiliation{Department of Physics, Graduate School of Science, The University of Tokyo, 7-3-1 Hongo, Bunkyo, Tokyo 113-0033, Japan}
\email{yana@icrr.u-tokyo.ac.jp}

\author[orcid=0000-0002-4225-4477]{Makoto Ando}
\affiliation{Institute for Cosmic Ray Research, The University of Tokyo, 5-1-5 Kashiwanoha, Kashiwa, Chiba 277-8582, Japan}
\email{mando@icrr.u-tokyo.ac.jp}

\author[orcid=0000-0002-6047-430X]{Yuichi Harikane}
\affiliation{Institute for Cosmic Ray Research, The University of Tokyo, 5-1-5 Kashiwanoha, Kashiwa, Chiba 277-8582, Japan}
\email{hari@icrr.u-tokyo.ac.jp}

\author[orcid=0009-0004-4332-9225]{Tomokazu Kiyota}
\affiliation{National Astronomical Observatory of Japan, 2-21-1 Osawa, Mitaka, Tokyo 181-8588, Japan}
\affiliation{Astronomical Science Program, Graduate Institute for Advanced Studies, SOKENDAI, 2-21-1 Osawa, Mitaka, Tokyo 181-8588, Japan}
\email{tomokazu.kiyota@grad.nao.ac.jp}

\author[orcid=0009-0000-1999-5472]{Minami Nakane}
\affiliation{Institute for Cosmic Ray Research, The University of Tokyo, 5-1-5 Kashiwanoha, Kashiwa, Chiba 277-8582, Japan}
\affiliation{Department of Physics, Graduate School of Science, The University of Tokyo, 7-3-1 Hongo, Bunkyo, Tokyo 113-0033, Japan}
\email{nakanem@icrr.u-tokyo.ac.jp}

\author[orcid=0000-0001-9011-7605]{Yoshiaki Ono}
\affiliation{Institute for Cosmic Ray Research, The University of Tokyo, 5-1-5 Kashiwanoha, Kashiwa, Chiba 277-8582, Japan}
\email{ono@icrr.u-tokyo.ac.jp}

\author[orcid=0009-0005-2897-002X]{Yui Takeda}
\affiliation{National Astronomical Observatory of Japan, 2-21-1 Osawa, Mitaka, Tokyo 181-8588, Japan}
\affiliation{Astronomical Science Program, Graduate Institute for Advanced Studies, SOKENDAI, 2-21-1 Osawa, Mitaka, Tokyo 181-8588, Japan}
\email{yui.takeda@grad.nao.ac.jp}

\begin{abstract}
We present an initial demographic study of Ly$\alpha$ profiles in little red dots (LRDs) at $z=3$--9 using $R\sim1000$--4000 spectroscopy. Our sample consists of 8 LRDs observed in the VLT/MUSE Deep and Wide surveys and 23 LRDs observed with JWST/NIRSpec grating spectroscopy from JADES, CANUCS, and GO programs including SPURS. We identify Ly$\alpha$ emission in 5 MUSE LRDs and 9 JWST LRDs. Only two of them exhibit broad Ly$\alpha$ emission (FWHM $>1000\ \mathrm{km\ s^{-1}}$), both reported previously. The other Ly$\alpha$-emitting LRDs show narrow Ly$\alpha$ emission (FWHM $<1000\ \mathrm{km\ s^{-1}}$), with FWHMs mostly in the range 300--600 $\mathrm{km\ s^{-1}}$, comparable to or slightly larger than those of high-redshift star-forming galaxies (100--500 $\mathrm{km\ s^{-1}}$). We measure the fraction of broad Ly$\alpha$ emitters above a broad Ly$\alpha$ luminosity threshold of $L_{\mathrm{Ly}\alpha,\mathrm{broad}}=10^{42}\ \mathrm{erg\ s^{-1}}$, obtaining $0.10^{+0.12}_{-0.07}$ for LRDs, about five times higher than the $2\sigma$ upper limit of $<0.02$ for high-redshift star-forming galaxies.
Although we reproduce the broad Ly$\alpha$ component reported in a previous stacking analysis of eight LRDs, albeit with a large uncertainty, we find no evidence for broad Ly$\alpha$ emission in either the larger JWST stack, after excluding the two individually detected broad Ly$\alpha$ emitters, or the higher-resolution MUSE stack. These results suggest that broad Ly$\alpha$ emission is not ubiquitous among LRDs. Instead, LRDs with broad Ly$\alpha$ emission appear to represent a rare population that may correspond to a particular evolutionary stage, potentially associated with unusually strong outflows or other distinctive physical conditions.



\end{abstract}


\section{Introduction}
\label{sec:introduction}

JWST has revealed an abundant population of compact sources at $z\sim4$--9 with blue rest-frame UV continua and red rest-frame optical continua, producing characteristic V-shaped spectral energy distributions \citep{harikane23agn, kocevski23, labbe23, greene24, kokorev24, matthee24, akins25, kocevski25}. These little red dots (LRDs) are typically unresolved or only marginally resolved, and many show broad Balmer emission together with narrow forbidden lines, indicating accreting supermassive black holes (SMBHs) \citep{kokorev23, matthee24}. Their physical nature nevertheless remains debated, because LRDs lack many of the signatures expected for AGNs. Almost no LRDs are individually detected in deep X-ray imaging, and stacking analyses return non-detections or at most marginal signals, with luminosities roughly an order of magnitude below the expectation from the X-ray--H$\alpha$ relation of typical unobscured AGNs \citep{ananna24, yue24, sacchi25, brazzini26}. High-ionization UV lines such as N~\textsc{v}$\lambda1240$, C~\textsc{iv}$\lambda1549$, and He~\textsc{ii}$\lambda1640$ are likewise weak or undetected, indicating an ionizing continuum considerably softer than that of standard quasars \citep{bingjiewang26, zucchi26}. Sub-millimeter observations also yield non-detections, both individually and in stacks, placing dust masses and infrared luminosities well below those required if the red continuum were produced by heavy dust obscuration \citep{casey25, setton25}.

In addition, many LRDs show a sharp continuum break near the Balmer limit, sometimes accompanied by H$\alpha$ or H$\beta$ absorption \citep{degraaff25cliff, inayoshimaiolino25, deugenio26abs}. Gas at $n_{\mathrm{H}}\sim10^{9}$--$10^{11}\ \mathrm{cm^{-3}}$ can collisionally populate the hydrogen $n=2$ state and generate both the break and Balmer absorption without a dominant stellar contribution \citep{inayoshimaiolino25}. Dense-cocoon and black-hole-envelope models accordingly place a rapidly accreting SMBH inside optically thick gas that reprocesses the ionizing radiation, broadens emission lines through electron scattering, and attenuates X-rays \citep{naidu25, naidu26, rusakov26, sneppen26, takasao26}. While observations and theoretical models increasingly support dense gas around LRD nuclei, its geometry, covering factor, and inflow/outflow configuration remain under debate.

Ly$\alpha$ provides a complementary probe of this dense-gas picture. Because it is a resonance line arising from ground-state neutral hydrogen, Ly$\alpha$ has a much larger effective opacity than the Balmer lines and is therefore particularly sensitive to the geometry and kinematics of gas surrounding an LRD \citep{dijkstra14}. Ly$\alpha$ emission has been identified in many LRDs, but whether it is powered predominantly by star formation in the host galaxy or by the central AGN remains unresolved. On the one hand, analyses of JWST/NIRSpec prism spectra have found Ly$\alpha$ equivalent-width distributions or stacked equivalent widths broadly consistent with those of star-forming galaxies \citep{ando26, asada26, zhiyuanji26lya}. VLT/MUSE observations of A2744-45924 also revealed a narrow ($\mathrm{FWHM}=270\pm15\ \mathrm{km\ s^{-1}}$) Ly$\alpha$ halo whose spatial and spectral properties favor an origin in host-galaxy star formation \citep{torralba26}. On the other hand, CANUCS-LRD-z8.6 and Abell2744-QSO1 exhibit broad Ly$\alpha$ emission with $\mathrm{FWHM}>1000\ \mathrm{km\ s^{-1}}$, pointing to an AGN origin \citep{xji26, morishita26, tang26}. Such emission may escape if the dense gas contains holes or forms a clumpy, porous envelope, with Ly$\alpha$ photons propagating through resonant scattering \citep{xji26, tang26}. Finally, \citet{geris26} reported a broad Ly$\alpha$ component in a stack of eight LRDs, raising the possibility that broad, AGN-related Ly$\alpha$ is present in more LRDs than individual detections imply, although this interpretation remains tentative.

This paper is the fourth in the Archival and Theoretical study of LRDs with AGN comparison across Surveys (ATLAS) series, which investigates the LRD population statistically using archival spectroscopy and interprets it through systematic comparisons with known AGN populations. The first three papers of the series examined the scaling relations between bolometric luminosity and broad H$\alpha$ and H$\beta$ luminosities \citep{yanagisawa26}, the incidence and kinematics of Balmer-line absorption \citep{yanagisawa26b}, and the hydrogen line ratios and dust content of LRDs \citep{kiyota26}. In this paper, we characterize Ly$\alpha$ properties in a statistical sample of LRDs to determine the prevalence of broad Ly$\alpha$ emission and assess the AGN contribution. We use VLT/MUSE and JWST/NIRSpec grating spectra for this purpose. Their spectral resolutions of $R\sim1000$--4000 enable us to resolve Ly$\alpha$ line profiles that cannot be measured with the $R\sim100$ NIRSpec prism spectra used in many previous studies. The paper is organized as follows. Section~\ref{sec:data_sample} describes the data and LRD sample selection. Section~\ref{sec:measurements} presents the measurements of Ly$\alpha$ properties. Section~\ref{sec:stacked_spectra} presents the stacked-spectrum analysis. Section~\ref{sec:results} presents the broad Ly$\alpha$ fraction and Ly$\alpha$ FWHM distribution, Section~\ref{sec:discussion} discusses their implications, and Section~\ref{sec:summary} summarizes our findings.

All magnitudes are in the AB system \citep{oke83}.
Throughout this work, we adopt the Planck 2018 cosmology with $H_0=67.66\ \mathrm{km\ s^{-1}\ Mpc^{-1}}$ and $\Omega_{\mathrm{m}}=0.30966$ \citep{planck20}.

\section{Data and Sample}
\label{sec:data_sample}
\subsection{Little Red Dots}
\label{sec:lrd_sample}
\subsubsection{Photometric Selection}
\label{sec:lrd_photometric_selection}
We select LRDs using JWST/NIRCam imaging and photometry in the GOODS-S, GOODS-N, EGS, Abell 2744, Abell 370, MACS 0416, MACS 0417, MACS 1149, and MACS 1423 fields.

For GOODS-S and GOODS-N, we use the JADES DR5 photometric catalogs and reduced NIRCam imaging \citep{johnson26, robertson26}.
The imaging data included in JADES DR5 were obtained through the JADES (GTO 1180, 1181, 1210, 1286, and 1287, PIs: Eisenstein \& Luetzgendorf; \citealt{eisenstein26}), PEARLS (GTO 1176, PI: Windhorst; \citealt{windhorst23}), GTO 1264 (PI: Colina), MIDIS (GTO 1283 and GO 6511, PI: Ostlin; \citealt{perezgonzalez24, ostlin25, perezgonzalez25}), FRESCO (GO 1895, PI: Oesch; \citealt{oesch23}), NGDEEP (GO 2079, PIs: Finkelstein, Papovich, \& Pirzkal; \citealt{bagley24}), GO 2198 (PIs: Barrufet \& Oesch), PANORAMIC (GO 2514, PIs: Williams \& Oesch; \citealt{williams25}), GO 2516 (PIs: Hodge \& da Cunha), JOF (GO 3215 and GTO 4540, PIs: Eisenstein \& Maiolino; \citealt{eisenstein25}), CONGRESS (GO 3577, PIs: Egami \& Sun), BEACON (GO 3990, PIs: Morishita, Mason, Treu, \& Trenti; \citealt{morishita25}), GO 4762 (PIs: Fujimoto \& Brammer), POPPIES (GO 5398, PIs: Kartaltepe \& Rafelski), OASIS (GO 5997, PIs: Looser \& D'Eugenio), SAPPHIRES (GO 6434, PIs: Egami, Fan, Sun, Wang, \& Yang; \citealt{sun25}), and DDT 6541 (PI: Egami) programs.

For EGS, we use the CEERS DR1 photometric catalog and reduced NIRCam imaging \citep{bagley23, finkelstein25, cox26}.
The imaging data included in CEERS DR1 were obtained through the CEERS (ERS 1345, PI: Finkelstein; \citealt{finkelstein25}) and DDT 2750 (PI: Arrabal Haro; \citealt{arrabalharo23}) programs.

For Abell 2744, we use the UNCOVER DR3 photometric catalog and reduced NIRCam imaging \citep{suess24, weaver24}.
The imaging data included in UNCOVER DR3 were obtained through the UNCOVER (GO 2561, PIs: Labbe \& Bezanson; \citealt{bezanson24}), GLASS (ERS 1324 and DDT 2756, PIs: Treu \& Chen; \citealt{treu22}), BEACON (GO 3990, PI: Morishita; \citealt{morishita25}), ALT (GO 3516, PIs: Matthee \& Naidu; \citealt{naidu24}), MAGNIF (GO 2883, PI: Sun; \citealt{fu25}), GO 3538 (PI: Iani), and PEARLS (GTO 1176, PI: Windhorst; \citealt{windhorst23}) programs.

For Abell 370, MACS 0416, MACS 0417, MACS 1149, and MACS 1423, we use the CANUCS/Technicolor DR1 photometric catalogs and reduced NIRCam imaging \citep{sarrouh26}.
The imaging data included in CANUCS/Technicolor DR1 were obtained through CANUCS (GTO 1208, PI: Willott) for all five cluster fields and Technicolor (GO 3362, PI: Muzzin) for the Abell 370, MACS 0416, and MACS 1149 fields.

From these catalogs we use the fluxes measured in $0.3''$-diameter apertures, except for Abell 2744, where we use the $0.32''$-diameter aperture fluxes.

We adopt the color and compactness thresholds presented by \citet{kokorev24} to select LRDs.
Following \citet{kokorev24}, we first require positive catalog fluxes in F115W, F150W, F200W, F277W, F356W, and F444W, an F444W detection significance of $>14\sigma$, and $m_{\mathrm{F444W}}<27.7$ mag.
Candidates must then satisfy either
\begin{displaymath}
\begin{array}{rcl}
m_{\mathrm{F115W}}-m_{\mathrm{F150W}} & < & 0.8,\\
m_{\mathrm{F200W}}-m_{\mathrm{F277W}} & > & 0.7,\\
m_{\mathrm{F200W}}-m_{\mathrm{F356W}} & > & 1.0,
\end{array}
\end{displaymath}
or
\begin{displaymath}
\begin{array}{rcl}
m_{\mathrm{F150W}}-m_{\mathrm{F200W}} & < & 0.8,\\
m_{\mathrm{F277W}}-m_{\mathrm{F356W}} & > & 0.6,\\
m_{\mathrm{F277W}}-m_{\mathrm{F444W}} & > & 0.7.
\end{array}
\end{displaymath}
We additionally require $m_{\mathrm{F115W}}-m_{\mathrm{F200W}}>-0.5$ to reduce contamination from brown dwarfs.
Finally, we measure F444W aperture fluxes directly from the imaging and require $f_{\mathrm{F444W}}(0.4''\ \mathrm{diameter})/f_{\mathrm{F444W}}(0.2''\ \mathrm{diameter})<1.7$ to select sources with a high central flux concentration.

The resulting photometrically selected sample contains 212 LRDs in GOODS-S, 119 in GOODS-N, 83 in EGS, 35 in Abell 2744, 12 in Abell 370, 11 in MACS 0416, 5 in MACS 0417, 14 in MACS 1149, and 7 in MACS 1423.

Because LRDs at $z>8$ are difficult to select efficiently using the color criteria above, we supplement the photometrically selected sample with five spectroscopically confirmed LRDs from the literature, which are C3PO-45290 at $z=8.3536$ \citep{papovich26}, SAPPHIRES-8760 at $z=8.479$ \citep{fudamoto25}, UNCOVER-20466 at $z=8.502$ \citep{kokorev23}, CANUCS-LRD-z8.6 at $z=8.6319$ \citep{tripodi25}, and CAPERS-LRD-z9 at $z=9.288$ \citep{taylor25}.
Of these five sources, only C3PO-45290 and CANUCS-LRD-z8.6 are included in the final spectroscopic sample described in Section~\ref{sec:lrd_spectroscopic_selection}, because they are the only ones with NIRSpec grating spectroscopy covering Ly$\alpha$ \citep{papovich26, morishita26}.

\subsubsection{Spectroscopic Sample Selection and Data}
\label{sec:lrd_spectroscopic_selection}
From the sources selected in Section~\ref{sec:lrd_photometric_selection}, we select those with systemic redshifts determined from JWST/NIRSpec spectroscopy using rest-frame optical emission lines, including the Balmer lines and [O~\textsc{iii}].
To obtain data with spectral resolving powers of $R\sim1000$--$4000$ for investigating their Ly$\alpha$ line profiles, we further require VLT/MUSE or JWST/NIRSpec G140M or G140H spectroscopy in which Ly$\alpha$ falls within the observed wavelength coverage.

For the MUSE sample, we examine all photometrically selected LRDs within the footprints of the MUSE-Deep\footnote{\url{https://amused.univ-lyon1.fr/project/UDF/}} and MUSE-Wide\footnote{\url{https://musewide.aip.de/}} surveys, including both Ly$\alpha$ detections and non-detections.
By crossmatching these sources with the public emission-line catalogs released by the MUSE teams \citep{urrutia19, bacon23}, we identify four Ly$\alpha$-emitting LRDs in MUSE-Deep and one in MUSE-Wide.
One of the four MUSE-Deep sources, GS-204851, is also covered by MUSE-Wide.
For this source, we use the deeper MUSE-Deep data.
For these five Ly$\alpha$-detected sources, we use the reduced data and one-dimensional spectra extracted and released by the respective MUSE teams.
These spectra are extracted with spatial weights based on the detected Ly$\alpha$ emission or source morphology, rather than with fixed apertures.
The weighted regions are typically about $1.5''$ across.
We additionally include three LRDs without Ly$\alpha$ detections that lie within the MUSE-Deep or MUSE-Wide datacube footprints, yielding a total of eight MUSE LRDs.
For these non-detections, we perform PSF-weighted extraction of one-dimensional spectra from the MUSE datacubes at the source positions measured from NIRCam imaging.

\begin{figure}[t]
\centering
\includegraphics[width=\columnwidth]{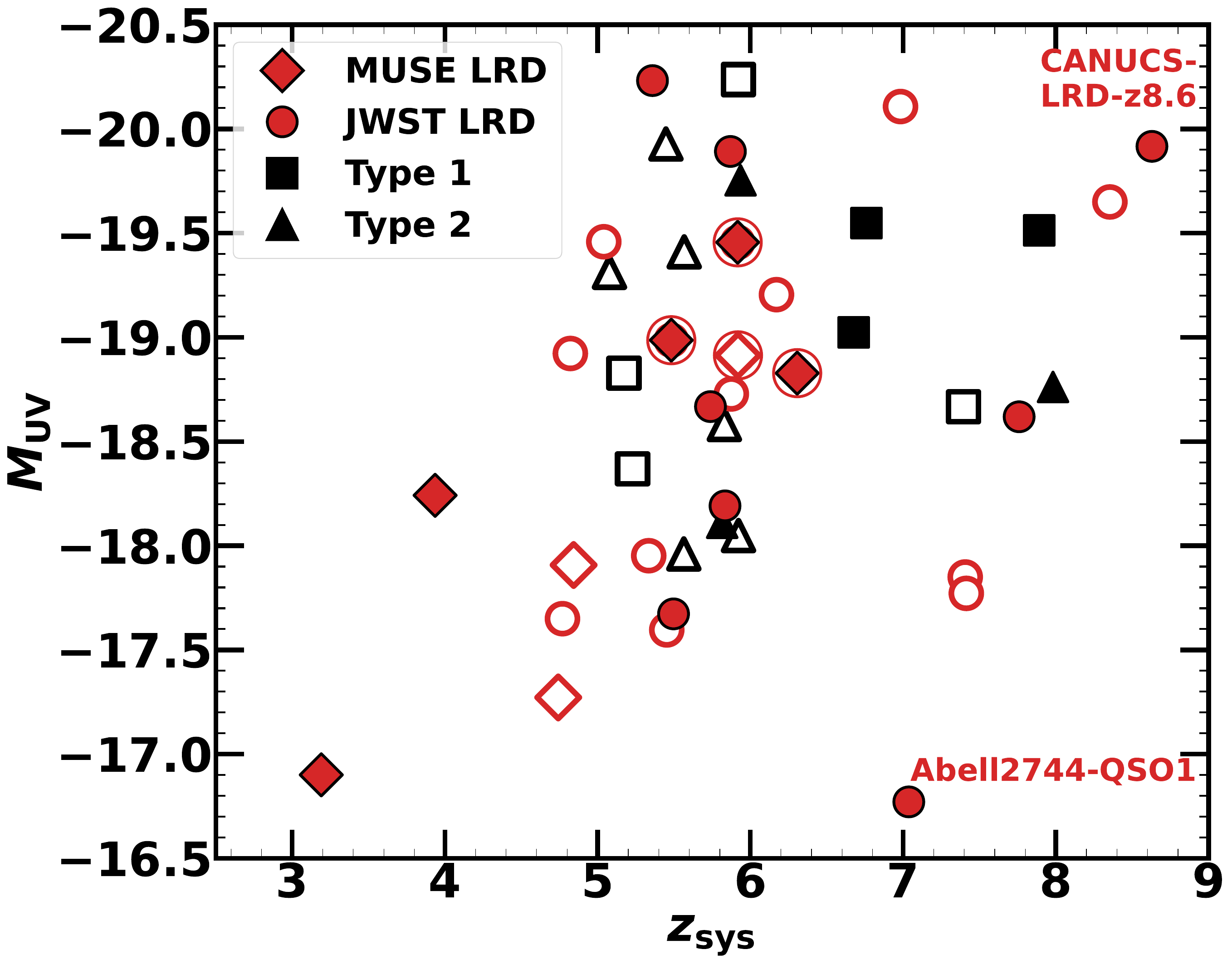}
\caption{Systemic-redshift and rest-frame UV absolute-magnitude distributions of the LRD and AGN samples. The $M_{\mathrm{UV}}$ values are corrected for gravitational lensing magnification. The symbols indicate MUSE LRDs, JWST LRDs, Type 1 AGNs, and Type 2 AGNs, as labeled. The filled and open symbols indicate sources with Ly$\alpha$ detections at S/N$>3$ and Ly$\alpha$ non-detections, respectively. The four LRDs included in both the MUSE and JWST samples are enclosed by red circles.}
\label{fig:z_muv}
\end{figure}

For the NIRSpec sample, we crossmatch the selected LRDs with observations from JADES (GTO 1180, 1181, 1210, 1286, and 1287, PIs: Eisenstein, Luetzgendorf, \& Isaak; \citealt{eisenstein26}), GO 4287 (PIs: Mason \& Stark; \citealt{whitler26}), CANUCS (GTO 4552, PI: Stiavelli; \citealt{morishita26}), DREAMS (GO 4750, PI: Nakajima; \citealt{nakajima26}), C3PO (GO 5943, PIs: Papovich, Hutchison, \& Hu; \citealt{papovich26}), DIVER (GO 8018, PI: Lin), and SPURS (GO 9214, PIs: Mason \& Stark; \citealt{chen26}).
For the G140M or G140H observations, the nominal wavelength ranges covered with the F070LP and F100LP filters are 0.70--1.27 and 0.97--1.89~$\mu$m, respectively.
We select sources whose Ly$\alpha$ wavelength at the systemic redshift falls within these wavelength ranges, irrespective of whether Ly$\alpha$ is detected.
JADES, CANUCS, DREAMS, and DIVER employ F070LP, so this requirement selects sources at $z>4.758$, whereas GO 4287, C3PO, and SPURS employ F100LP, selecting sources at $z>6.979$.
We visually inspect the spectra and exclude sources with data-quality issues, such as contamination from nearby objects.
This selection yields 23 LRDs, including both Ly$\alpha$ detections and non-detections.
Four sources, GS-9698, GS-13704, GS-204851, and GS-210600, are included in both the MUSE and NIRSpec samples.
To determine the systemic redshifts of CANUCS-LRD-z8.6 and GS-200576, we use NIRSpec prism spectra from CANUCS (GTO 1208, PI: Willott; \citealt{sarrouh26}) and GO 8060 (PIs: Egami, Maiolino, \& Rest), respectively.
To investigate the rest-frame optical continuum of Abell2744-QSO1 in Section~\ref{sec:discussion}, we use the NIRSpec prism spectrum obtained by UNCOVER (GO 2561, PIs: Labbe \& Bezanson; \citealt{bezanson24}).
When a spectrum is available in version 4.4 of the DAWN JWST Archive (DJA)\footnote{\url{https://dawn-cph.github.io/dja/}}, we use the publicly released spectrum reduced with \texttt{msaexp} \citep{brammer23, heintz24, degraaff25, valentino25, pollock26}.
For sources without a public DJA spectrum, we reduce the NIRSpec data ourselves using \texttt{msaexp} with the Calibration Reference Data System (CRDS) context file \texttt{jwst\_1322.pmap}.
The resulting spectroscopic LRD sample and corresponding program assignments are listed in Table~\ref{tab:lrd_spectroscopic_sample}.
The distributions of systemic redshift and rest-frame UV absolute magnitude for the LRD sample are shown in Figure~\ref{fig:z_muv}.

\newcommand{\lrdtableline}{\noalign{\begingroup\color{black!25}\hrule height 0.2pt\endgroup}}
\begingroup
\setlength{\tabcolsep}{2pt}
\begin{deluxetable*}{llccccccccc}
\tabletypesize{\scriptsize}
\tablecaption{LRD Spectroscopic Sample\label{tab:lrd_spectroscopic_sample}}
\tablehead{
\colhead{Field} & \colhead{ID} & \colhead{PID} & \colhead{R.A.} & \colhead{Decl.} & \colhead{$z_{\mathrm{sys}}$} & \colhead{$M_{\mathrm{UV}}$} & \colhead{$F_{\mathrm{Ly}\alpha}$} & \colhead{$\Delta v_{\mathrm{Ly}\alpha}$} & \colhead{$\mathrm{FWHM}_{\mathrm{Ly}\alpha}$} & \colhead{Reference} \\
\colhead{(1)} & \colhead{(2)} & \colhead{(3)} & \colhead{(4)} & \colhead{(5)} & \colhead{(6)} & \colhead{(7)} & \colhead{(8)} & \colhead{(9)} & \colhead{(10)} & \colhead{(11)} \\
\colhead{} & \colhead{} & \colhead{} & \colhead{(deg)} & \colhead{(deg)} & \colhead{} & \colhead{} & \colhead{\shortstack{($10^{-18}\,\mathrm{erg\,s^{-1}}$\\$\mathrm{cm^{-2}}$)}} & \colhead{($\mathrm{km\,s^{-1}}$)} & \colhead{($\mathrm{km\,s^{-1}}$)} & \colhead{}
}
\startdata
GOODS-S & GS-5356 & 1180 & 53.179855 & -27.808283 & 5.8338 & -18.2 & $3.50^{+0.97}_{-1.1}$ & $<176$ & $557^{+100}_{-300}$ & GJ25 \\
\lrdtableline
GOODS-S & GS-9698 & 1180 & 53.140780 & -27.802180 & 5.9168 & -19.5 & $<5.9$ &  &  & GJ25 \\
 &  & DEEP &  &  &  &  & $6.25\pm0.46$ & $120^{+11}_{-13}$ & $343^{+26}_{-33}$ & PR25 \\
\lrdtableline
GOODS-S & GS-11786 & 1287 & 53.119072 & -27.892553 & 7.4065 & -17.9 & $<0.955$ &  &  & PR25 \\
\lrdtableline
GOODS-S & GS-12402 & WIDE & 53.132664 & -27.765488 & 3.1904 & -16.9 & $23^{+3}_{-4}$ & $<57.3$ & $219^{+60}_{-80}$ & AG25 \\
\lrdtableline
GOODS-S & GS-13329 & DEEP & 53.139034 & -27.784441 & 3.9357 & -18.2 & $6.13\pm0.61$ & $194^{+20}_{-40}$ & $484^{+50}_{-64}$ & AG25, PR25 \\
 &  &  &  &  &  &  &  &  &  & IJ26 \\
\lrdtableline
GOODS-S & GS-13418 & 1286 & 53.107220 & -27.890602 & 6.1707 & -19.2 & $<3.60$ &  &  & PR25 \\
\lrdtableline
GOODS-S & GS-13704 & 1210 & 53.126537 & -27.818089 & 5.9187 & -18.9 & $<2.8$ &  &  & PR25 \\
 &  & WIDE &  &  &  &  & $<5.52$ &  &  & IJ26 \\
\lrdtableline
GOODS-S & GS-38562 & 1286 & 53.135860 & -27.871645 & 4.8213 & -18.9 & $<9.11$ &  &  & AG25, IJ26 \\
\lrdtableline
GOODS-S & GS-73690 & 1286 & 53.060543 & -27.848398 & 5.4968 & -17.7 & $14.2\pm1.6$ & $102^{+40}_{-90}$ & $416^{+50}_{-60}$ & GJ25 \\
 &  &  &  &  &  &  &  &  &  & PR25 \\
\lrdtableline
GOODS-S & GS-171875 & 1286 & 53.063156 & -27.873410 & 5.7386 & -18.7 & $6.0^{+1.2}_{-1.3}$ & $<179$ & $581^{+100}_{-200}$ & GJ25 \\
\lrdtableline
GOODS-S & GS-172975 & WIDE & 53.087730 & -27.871237 & 4.7420 & -17.3 & $<4.08$ &  &  & GJ25, PR25 \\
 &  &  &  &  &  &  &  &  &  & IJ26 \\
\lrdtableline
GOODS-S & GS-200576 & DEEP & 53.154770 & -27.806522 & 4.8427 & -17.9 & $<1.42$ &  &  & PR25 \\
\lrdtableline
GOODS-S & GS-204851 & 1286 & 53.138590 & -27.790253 & 5.4821 & -19.0 & $<8.7$ &  &  & AG25, GJ25 \\
 &  & \raisebox{-3pt}[0pt][0pt]{DEEP} &  &  &  &  & $40.20\pm0.77~(\mathrm{red})$ & $20\pm3$ & $286.9\pm8.3$ & PR25, IJ26 \\
 &  &  &  &  &  &  & $2.74^{+0.70}_{-0.47}~(\mathrm{blue})$ & $-327^{+41}_{-12}$ & $113^{+20}_{-60}$ &  \\
\lrdtableline
GOODS-S & GS-210600 & 1286 & 53.166115 & -27.772040 & 6.3064 & -18.8 & $5.77\pm0.64$ & $<143$ & $416^{+50}_{-60}$ & GJ25, PR25 \\
 &  & DEEP &  &  &  &  & $8.63\pm0.39$ & $24\pm5$ & $307^{+14}_{-16}$ & IJ26 \\
\lrdtableline
GOODS-N & GN-4685 & 1181 & 189.096300 & 62.239150 & 7.4136 & -17.8 & $<2.27$ &  &  & AG25, PR25 \\
 &  &  &  &  &  &  &  &  &  & IJ26 \\
\lrdtableline
GOODS-N & GN-38147 & 1181 & 189.270680 & 62.148426 & 5.8690 & -19.9 & $7.6^{+2.0}_{-2.2}$ & $<346$ & $915^{+200}_{-300}$ & PR25 \\
\lrdtableline
GOODS-N & GN-61888 & 1181 & 189.168010 & 62.217022 & 5.8752 & -18.7 & $<2.46$ &  &  & PR25, IJ26, XL26 \\
\lrdtableline
GOODS-N & GN-68797 & 1181 & 189.229170 & 62.146190 & 5.0402 & -19.5 & $<10.3$ &  &  & AG25, PR25 \\
\lrdtableline
GOODS-N & GN-1020514 & 8018 & 189.179300 & 62.292538 & 5.3590 & -20.2 & $23^{+4}_{-9}~(\mathrm{narrow})$ & $344\pm16$ & $224^{+100}_{-100}$ & XL26 \\
 &  &  &  &  &  &  & $29^{+9}_{-4}~(\mathrm{broad})$ & $<98.1$ & $841^{+31}_{-200}$ &  \\
\lrdtableline
GOODS-N & GN-1066100 & 8018 & 189.227200 & 62.234142 & 4.7696 & -17.6 & $<9.82$ &  &  & XL26 \\
\lrdtableline
EGS & CEERS-20 & 4287 & 214.830658 & 52.887775 & 7.7595 & -18.6 & $1.37^{+0.36}_{-0.42}$ & $229^{+30}_{-40}$ & $270.0^{+3.4}_{-200}$ & SF23 \\
\lrdtableline
EGS & CEERS-7902 & 9214 & 214.983032 & 52.956001 & 6.9829 & -20.1 & $<0.357$ &  &  & MT25 \\
\lrdtableline
EGS & C3PO-45290 & 5943 & 214.876144 & 52.880825 & 8.3538 & -19.6 & $<1.08$ &  &  & CP26 \\
\lrdtableline
Abell 2744 & Abell2744- & 9214 & 3.583535 & -30.396679 & 7.0372 & -16.8 & $5.29\pm0.27$ & $<136$ & $1623^{+90}_{-100}$ & LF24, AG25 \\
 & QSO1 &  &  &  &  &  &  &  &  & XJ26, MT26 \\
\lrdtableline
MACS 0416 & DREAMS- & 4750 & 64.052488 & -24.081448 & 5.3343 & -18.0 & $<0.859$ &  &  & SW26, TW \\
 & 60003 &  &  &  &  &  &  &  &  & \\
\lrdtableline
MACS 0416 & DREAMS- & 4750 & 64.076870 & -24.103464 & 5.4524 & -17.6 & $<17.3$ &  &  & TW \\
 & 60007 &  &  &  &  &  &  &  &  & \\
\lrdtableline
MACS 1149 & CANUCS-LRD- & 4552 & 177.390930 & 22.349767 & 8.6291 & -19.9 & $2.50\pm0.65$ & $<525$ & $1922^{+500}_{-500}$ & RT25 \\
 & z8.6 &  &  &  &  &  &  &  &  & TM26 \\
\enddata
\tablecomments{(1) Field. (2) Source ID. (3) Program identifier. DEEP and WIDE denote MUSE-DEEP and MUSE-WIDE, respectively, and the JWST/NIRSpec program ID is listed otherwise. (4), (5) R.A. and Decl. (6) Systemic redshift. (7) Rest-frame UV absolute magnitude corrected for gravitational lensing magnification. (8) Ly$\alpha$ flux, not corrected for gravitational lensing magnification. (9) Ly$\alpha$ velocity offset. (10) Ly$\alpha$ FWHM. (11) Reference. GJ25: \citet{jones25}. PR25: \citet{rinaldi25}. AG25: \citet{degraaff25}. IJ26: \citet{juodzbalis26}. XL26: \citet{lin26}. SF23: \citet{fujimoto23}. MT25: \citet{tang25}. CP26: \citet{papovich26}. LF24: \citet{furtak24}. XJ26: \citet{xji26}. MT26: \citet{tang26}. SW26: \citet{withers26}. TW: This Work. RT25: \citet{tripodi25}. TM26: \citet{morishita26}. For the fitted Ly$\alpha$ quantities in columns (8)--(10), central values are posterior means, and the uncertainties are 68\% credible intervals. Upper limits on $F_{\mathrm{Ly}\alpha}$ and $\Delta v_{\mathrm{Ly}\alpha}$ are the 95\% and 68\% upper limits, respectively. For GS-204851 and GN-1020514, which are fit with two components, the two successive rows list the red and blue components, respectively, for GS-204851 and the narrow and broad components, respectively, for GN-1020514.}
\end{deluxetable*}
\endgroup

\subsection{Non-LRD Type 1 and Type 2 AGNs}
\label{sec:non_lrd_agn}

To construct the non-LRD AGN comparison samples, we adopt the JADES spectroscopic broad-line (Type 1) AGN catalog of \citet{juodzbalis26} and the narrow-line (Type 2) AGN candidate catalog of \citet{scholtz25agn}.
For the Type 1 sample, we explicitly remove all sources included in the LRD sample defined in Section~\ref{sec:lrd_sample}.
We apply the same non-LRD requirement to the Type 2 sample.
Our Type 1 AGN sample is not defined as a sample of ``little blue dots'' (LBDs), a term used for compact broad-line AGNs with blue continua \citep{brazzini26, geris26}.
Because our exclusion is tied specifically to the LRD criteria adopted in Section~\ref{sec:lrd_sample}, the resulting Type 1 sample may include sources that would be classified as LRDs under alternative definitions, as well as broad-line AGNs that are not compact.
We further require an available JWST/NIRSpec G140M or G140H spectrum.
For these sources, we use the DJA spectra from JADES and JOF (GO 3215, PIs: Eisenstein \& Maiolino).
These criteria yield seven Type 1 AGNs and nine Type 2 AGNs.
The individual sources are listed in Tables~\ref{tab:type1_agn_sample} and \ref{tab:type2_agn_sample} in Appendix~\ref{sec:agn_sample_tables}.

\subsection{Star-Forming Galaxies}
\label{sec:sfg_sample}

To enable a fair comparison with the MUSE and NIRSpec LRD samples described in Section~\ref{sec:lrd_spectroscopic_selection}, we construct two complementary samples of star-forming galaxies (SFGs).

To construct the MUSE SFG comparison sample with measured systemic redshifts, we crossmatch the MUSE-Deep Ly$\alpha$ emitters with the DJA NIRSpec catalog described in Section~\ref{sec:lrd_spectroscopic_selection}.
We require a DJA quality grade of 3 and a redshift determined from a NIRSpec grating spectrum.
We first identify positional matches within $1.0''$ and then require the MUSE and NIRSpec redshifts to agree within $3000\ \mathrm{km\ s^{-1}}$.
Two MUSE sources have multiple distinct DJA objects satisfying these criteria.
For these two sources, we adopt the DJA object with the smallest angular separation.
We remove sources matched to the LRD sample defined in Section~\ref{sec:lrd_sample}, as well as sources included in the Type 1 or Type 2 AGN catalogs adopted in Section~\ref{sec:non_lrd_agn}.
The resulting MUSE comparison sample contains 124 SFGs with MUSE Ly$\alpha$ measurements and NIRSpec-based systemic redshifts.

For the NIRSpec comparison sample, we use the public JADES DR3 medium-resolution grating catalogs and spectra in GOODS-S and GOODS-N \citep{bunker24, deugenio25dr3}.
We require a spectroscopic redshift flag of A, indicating that the redshift is determined from emission lines in the grating spectra.
For each source, we use the G140M spectrum for Ly$\alpha$ and the G395M spectrum for H$\alpha$, requiring each spectrum to cover the corresponding line at the spectroscopic redshift.
The sources satisfying this line-coverage requirement span $4.79\leq z_{\mathrm{sys}}\leq6.99$.
This criterion requires coverage of both lines but does not demand a Ly$\alpha$ detection.
We then remove the LRDs and Type 1 and Type 2 AGNs identified in Sections~\ref{sec:lrd_sample} and \ref{sec:non_lrd_agn}.
The final JADES comparison sample contains 161 SFGs, with 64 in GOODS-S and 97 in GOODS-N.

\section{Measurements}
\label{sec:measurements}
\subsection{Systemic Redshift}
\label{sec:systemic_redshift}

Throughout this work, we use the H$\alpha$ redshift as the systemic redshift against which the Ly$\alpha$ velocity profile is measured.
For sources without H$\alpha$ coverage, we instead use the H$\beta$ redshift.
We fit the H$\alpha$+[N~\textsc{ii}] region in every available NIRSpec grating spectrum that covers these lines.
For each spectrum, we calculate a wavelength-dependent line-spread function (LSF) with \texttt{msafit} for a point source at its measured position within the MSA slit \citep{degraaff24msafit}.
When H$\alpha$ is covered by multiple gratings, we fit all of the spectra simultaneously with common physical line parameters while convolving the model for each spectrum with its corresponding LSF.

We fit the line profile in three successive stages.
First, we fit a single-component model in which H$\alpha$ is represented by one Gaussian.
This model also includes a linear continuum and the [N~\textsc{ii}] doublet, with H$\alpha$ and [N~\textsc{ii}] sharing the same redshift and intrinsic velocity dispersion.
We fix the [N~\textsc{ii}] $\lambda6549/\lambda6585$ flux ratio to 0.3399, calculated with \texttt{PyNeb} \citep{luridiana15}.
Second, we fit a single+exponential model that keeps the narrow H$\alpha$ component and adds a broad H$\alpha$ component with exponential wings.
The broad profile is a weighted sum of an intrinsic Gaussian and its convolution with a symmetric two-sided exponential profile \citep{rusakov26, scholtz26profiles}.
Finally, we fit a single+exponential+absorption model in which partial-covering Balmer absorption is applied to the continuum and broad H$\alpha$ component, while the narrow H$\alpha$ and [N~\textsc{ii}] components remain unabsorbed \citep{deugenio26abs}.

\begin{figure}[t]
\centering
\includegraphics[width=\columnwidth]{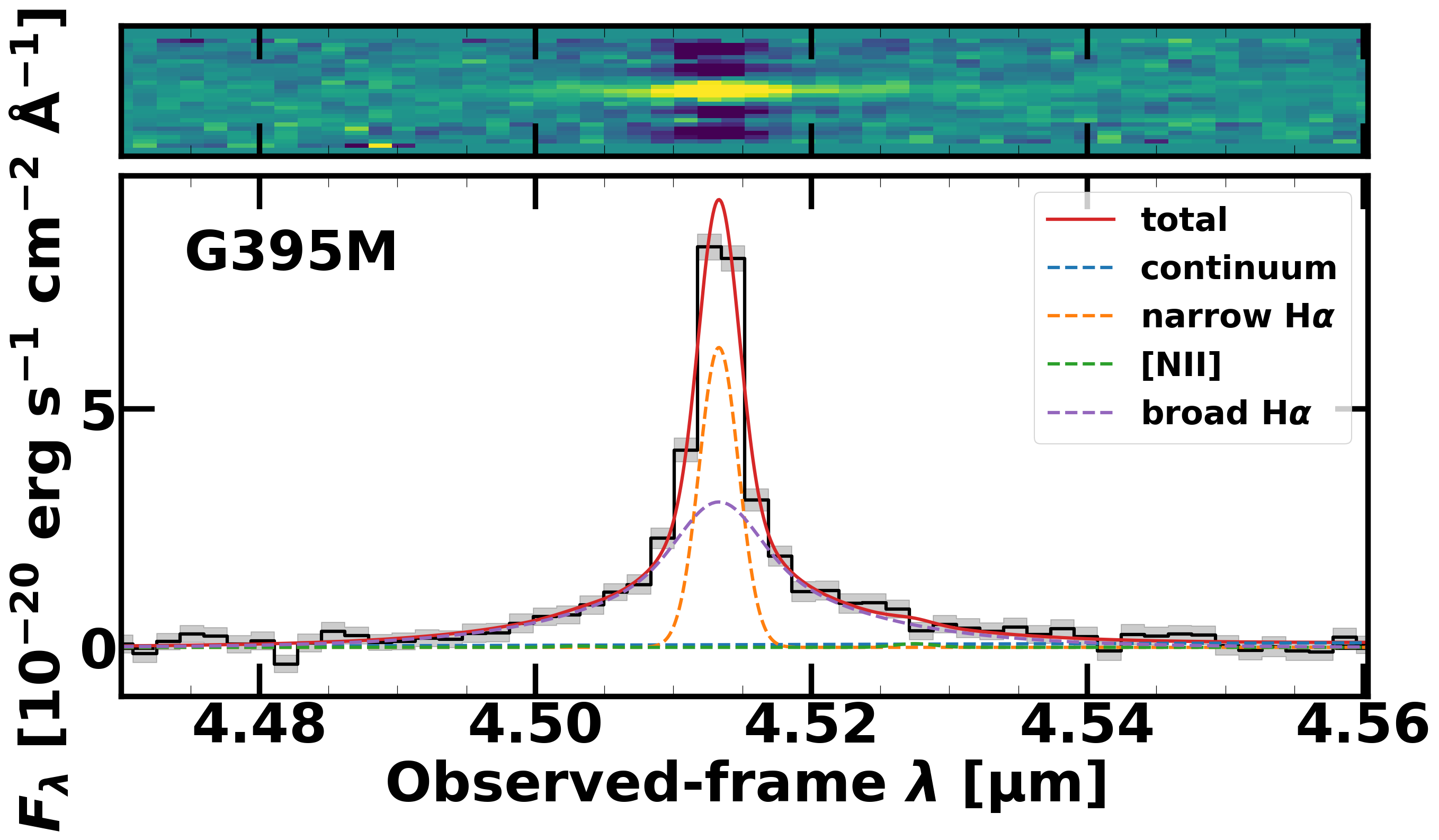}
\caption{Fit to the H$\alpha$+[N~\textsc{ii}] region of GN-61888 using the G395M spectrum. The two-dimensional spectrum is shown above its extracted one-dimensional spectrum. The black line and gray shading show the observed spectrum and its $1\sigma$ uncertainty. The red curve is the total model. The blue, orange, green, and purple dashed curves show the linear continuum, narrow H$\alpha$, [N~\textsc{ii}], and broad-component contributions, respectively. The single+exponential model is selected for this source.}
\label{fig:gn61888_balmer_fit}
\end{figure}

We sample the posterior distributions of all three models with \texttt{emcee} \citep{foremanmackey13}.
For model selection, we calculate the widely applicable information criterion (WAIC) \citep{watanabe10}.
We select the single+exponential model over the single-component model when $\Delta\mathrm{WAIC}=\mathrm{WAIC}_{\mathrm{single}}-\mathrm{WAIC}_{\mathrm{single+exp}}\geq11.8$ and the broad component has a signal-to-noise ratio of at least 3, following the criteria used by \citet{hviding25} and \citet{nakane26}.
If these requirements are satisfied, we select the absorption model when $\mathrm{WAIC}_{\mathrm{single+exp}}-\mathrm{WAIC}_{\mathrm{single+exp+abs}}\geq11.8$.
We adopt the maximum-posterior H$\alpha$ redshift from the selected model as the systemic redshift.
For sources without an H$\alpha$ grating spectrum that are fitted using H$\beta$, the low signal-to-noise ratio of H$\beta$ can result in a large uncertainty in the systemic redshift.
When the $1\sigma$ uncertainty in the systemic velocity derived from H$\beta$ exceeds $30\ \mathrm{km\ s^{-1}}$, we determine the systemic redshift by fitting [O~\textsc{iii}] $\lambda\lambda4960,5008$.
For GN-1066100 and CANUCS-LRD-z8.6, whose rest-frame optical grating spectra are unavailable, we determine the systemic redshifts from the prism spectra, fitting H$\beta$+[O~\textsc{iii}] $\lambda\lambda4960,5008$+H$\alpha$ for GN-1066100 and H$\beta$+[O~\textsc{iii}] $\lambda\lambda4960,5008$ for CANUCS-LRD-z8.6.
Figure~\ref{fig:gn61888_balmer_fit} shows an example G395M fit for GN-61888, for which the single+exponential model without Balmer absorption is selected.

\subsection{\texorpdfstring{$M_{\mathrm{UV}}$}{MUV} Measurements}
\label{sec:muv_measurements}

For all sources introduced in Section~\ref{sec:data_sample}, we measure $M_{\mathrm{UV}}$ using the photometric catalogs described in Section~\ref{sec:lrd_photometric_selection}, adopting JADES DR5 for GOODS-S and GOODS-N, CEERS DR1 for EGS, UNCOVER DR3 for Abell 2744, and CANUCS/Technicolor DR1 for MACS 0416 and MACS 1149 \citep{suess24, weaver24, johnson26, robertson26, sarrouh26, cox26}.
We primarily use the same NIRCam aperture-corrected catalog fluxes adopted for the LRD selection in Section~\ref{sec:lrd_photometric_selection}, selecting a broad band that contains rest-frame $1500\ \text{\AA}$ at the systemic redshift of each source.
For MUSE sources in GOODS-S at $z\sim3$--4, where rest-frame $1500\ \text{\AA}$ falls blueward of the usable NIRCam wavelength coverage, we instead use HST/ACS photometry from the Hubble Legacy Fields included in the JADES photometric catalog \citep{illingworth16hlf, whitaker19hlf}.
Before calculating $M_{\mathrm{UV}}$, we correct the fluxes for gravitational lensing magnification.
We adopt magnification factors of $\mu=7.29$ for Abell2744-QSO1 \citep{furtak24}, $\mu=1.85$ for CANUCS-LRD-z8.6 \citep{schuldt24, morishita26}, and $\mu=1.59$ and $1.15$ for DREAMS-60003 and DREAMS-60007, respectively, from the CANUCS photometric catalog \citep{sarrouh26}.

\subsection{Ly\texorpdfstring{$\alpha$}{alpha} Fitting}
\label{sec:lya_fitting}

\begin{figure*}[t]
\centering
\includegraphics[width=\textwidth]{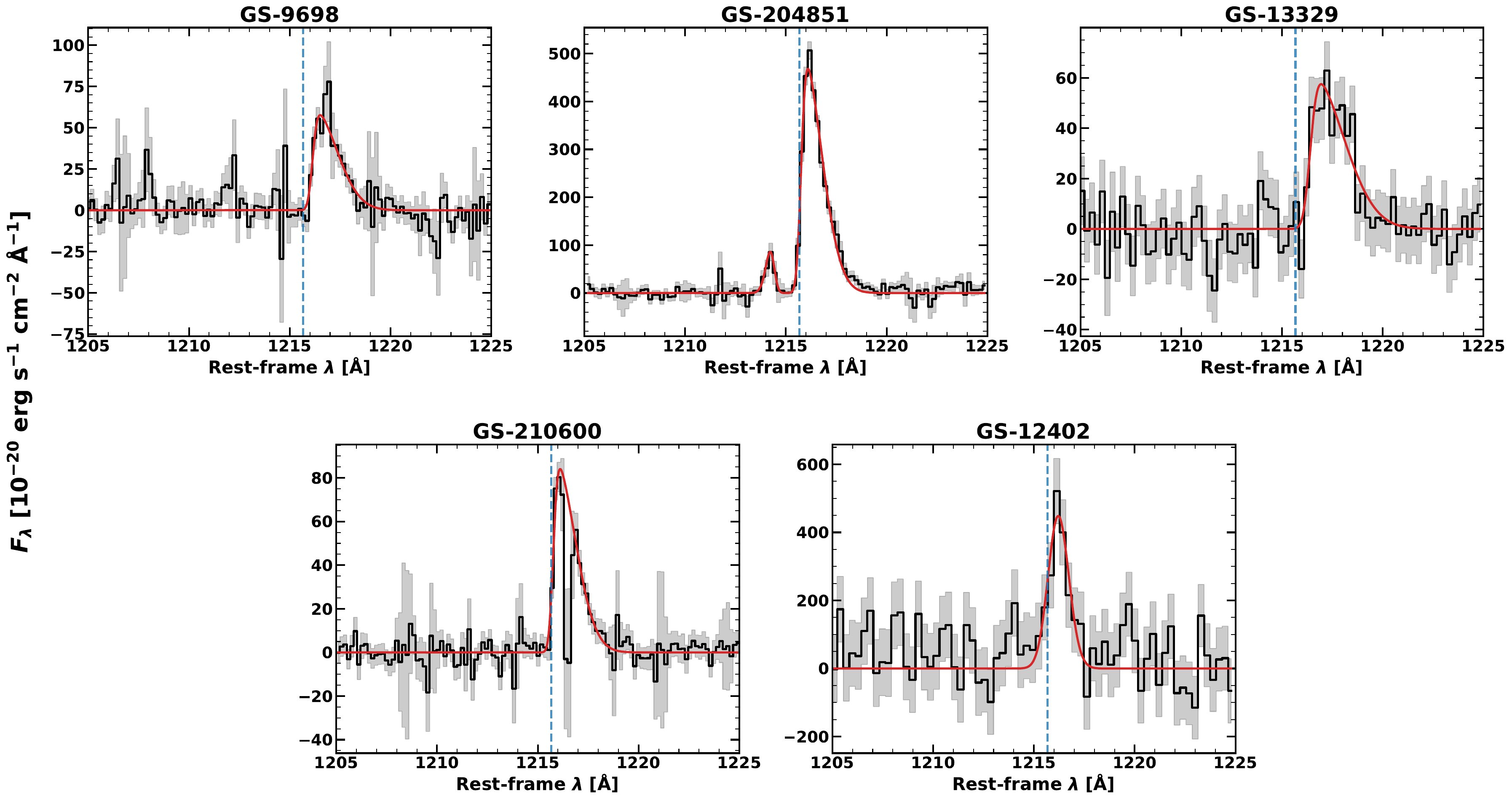}
\caption{MUSE Ly$\alpha$ spectra of the five Ly$\alpha$-detected LRDs and their best-fitting skewed-Gaussian models. The black lines and gray shading show the spectra and their $1\sigma$ uncertainties, respectively. The red curves show the best-fitting models, and the blue dashed lines indicate the Ly$\alpha$ wavelengths at the systemic redshifts. GS-204851 shows a blue Ly$\alpha$ peak, and its red and blue peaks are fitted with two skewed Gaussians. The sharp drop near rest-frame $1216.5$~\AA\ in GS-210600 is caused by oversubtraction of a sky line.}
\label{fig:muse_lrd_lya_spectra}
\end{figure*}

For the MUSE LRDs, we fit the Ly$\alpha$ profile with a skewed Gaussian, following the form used by \citet{bacon23}.
We use the systemic redshift determined in Section~\ref{sec:systemic_redshift} for the Ly$\alpha$ fitting.
The skewed-Gaussian model is
\begin{equation}
\begin{array}{rcl}
f(\lambda) & = & \frac{F_{\mathrm{Ly}\alpha}}{\sqrt{2\pi}\sigma_{\lambda}}\left[1+\mathrm{erf}\left(\frac{\gamma(\lambda-\lambda_0)}{\sqrt{2}\sigma_{\lambda}}\right)\right]\\
& & \times\exp\left[-\frac{(\lambda-\lambda_0)^2}{2\sigma_{\lambda}^2}\right],
\end{array}
\label{eq:muse_lya_model}
\end{equation}
where $F_{\mathrm{Ly}\alpha}$ is the integrated line flux, $\lambda_0=1215.67(1+z_{\mathrm{sys}})(1+\Delta v/c)$, $\sigma_{\lambda}=\sigma\lambda_0/c$, and $\gamma$ describes the asymmetry.
The parameters $z_{\mathrm{sys}}$, $\Delta v$, and $\sigma$ are the systemic redshift, the Ly$\alpha$ velocity offset relative to the systemic redshift, and the velocity dispersion, respectively.
The four free parameters of each skewed-Gaussian component are $F_{\mathrm{Ly}\alpha}$, $\Delta v$, $\sigma$, and $\gamma$.
We convolve the model with the instrumental LSF.
For the MUSE-Deep spectra, we use the instrumental velocity dispersion listed in the public emission-line catalog.
For the MUSE-Wide spectrum, we use the wavelength-dependent LSF from \citet{bacon23}, $\mathrm{FWHM}_{\mathrm{LSF}}=5.866\times10^{-8}\lambda^2-9.187\times10^{-4}\lambda+6.040$, where the FWHM and $\lambda$ are in \AA.
We impose $F_{\mathrm{Ly}\alpha}>0$, $10<2\sqrt{2\ln2}\,\sigma<2000\ \mathrm{km\ s^{-1}}$, $0<\gamma<10$, and $0<\Delta v<1000\ \mathrm{km\ s^{-1}}$.
GS-204851 has a reported blue Ly$\alpha$ component \citep{bacon23, ji26}, so we fit the red and blue components with two skewed Gaussians.
For the blue component, we impose $-1000<\Delta v<0\ \mathrm{km\ s^{-1}}$ and $-10<\gamma<0$, while keeping $F_{\mathrm{Ly}\alpha}>0$ and the same FWHM prior.
The MUSE Ly$\alpha$ spectra and their best-fitting models are shown in Figure~\ref{fig:muse_lrd_lya_spectra}.

For the NIRSpec sample, most sources are observed with G140M at $R\sim1000$, which makes it difficult to measure an asymmetric Ly$\alpha$ profile accurately.
We therefore use a Gaussian rather than a skewed-Gaussian model.
We use the G140M or G140H spectrum that covers Ly$\alpha$ and fit the rest-frame $1190$--$1240$~\AA\ range.
We model the Ly$\alpha$ emission line alone and do not include a continuum component.
To verify that omitting the continuum does not affect the grating-spectrum fit, we fit a power-law continuum to the prism data for the sources that have a prism spectrum from the same NIRSpec program and slit position.
For these sources, the inferred continuum is negligible compared with the uncertainties in the G140M or G140H spectrum.

We model the intrinsic Ly$\alpha$ profile as a Gaussian attenuated by the IGM:
\begin{equation}
\begin{array}{rcl}
f(\lambda) & = & \frac{F_{\mathrm{Ly}\alpha}}{\sqrt{2\pi}\sigma_{\lambda}}\exp\left\{-\frac{(\lambda-\lambda_0)^2}{2\sigma_{\lambda}^2}\right\}\\
& & \times\exp\left[-\tau_{\mathrm{IGM}}(\lambda,z_{\mathrm{sys}})\right],
\end{array}
\label{eq:lya_model}
\end{equation}
where $F_{\mathrm{Ly}\alpha}$ is the IGM-unattenuated intrinsic line flux, $\lambda_0=1215.67(1+z_{\mathrm{sys}})(1+\Delta v/c)$, $\sigma_{\lambda}=\sigma\lambda_0/c$, and $\tau_{\mathrm{IGM}}$ is the IGM optical depth following \citet{inoue14}.
The parameters $z_{\mathrm{sys}}$, $\Delta v$, and $\sigma$ have the same meanings as in Equation~(\ref{eq:muse_lya_model}).
We convolve the line profile with the wavelength-dependent LSF calculated with \texttt{msafit}.
The three free parameters of the single-Gaussian model are therefore $F_{\mathrm{Ly}\alpha}$, $\Delta v$, and $\sigma$.
The corresponding IGM-unattenuated intrinsic FWHM is defined as $\mathrm{FWHM}_{\mathrm{int}}=2\sqrt{2\ln2}\,\sigma$.

We first fit a single Gaussian with $0<\Delta v<1000\ \mathrm{km\ s^{-1}}$ and $10<\mathrm{FWHM}_{\mathrm{int}}<2000\ \mathrm{km\ s^{-1}}$, sampling the posterior with \texttt{emcee} \citep{foremanmackey13}.
For broad Ly$\alpha$ emitters, the posterior of $\sigma$ can reach the upper prior boundary of $2000/(2\sqrt{2\ln2})\ \mathrm{km\ s^{-1}}$, yielding only a lower limit on $\sigma$.
If the IGM-attenuated Ly$\alpha$ flux is detected at S/N$>3$ and the posterior of $\sigma$ has only this lower limit, we repeat the single-Gaussian fit with the expanded prior $10<\mathrm{FWHM}_{\mathrm{int}}<5000\ \mathrm{km\ s^{-1}}$.

For each source with an IGM-attenuated Ly$\alpha$ detection at S/N$>3$, we also fit a double-Gaussian model.
Each component has independent flux, velocity offset, and velocity dispersion, and the second component is required to be broader than the first.
For the broad component, we impose $10<\mathrm{FWHM}_{\mathrm{int}}<5000\ \mathrm{km\ s^{-1}}$.
We select the double-Gaussian model only when the broad component is detected at S/N$>3$ and $\Delta\mathrm{WAIC}=\mathrm{WAIC}_{\mathrm{single}}-\mathrm{WAIC}_{\mathrm{double}}>11.8$.
Otherwise, we adopt the single-Gaussian model.
From the selected model, we measure the Ly$\alpha$ velocity offset, the IGM-attenuated line flux, and the IGM-attenuated FWHM, $\mathrm{FWHM}_{\mathrm{Ly}\alpha}$.

\begin{figure*}[t]
\centering
\includegraphics[width=\textwidth]{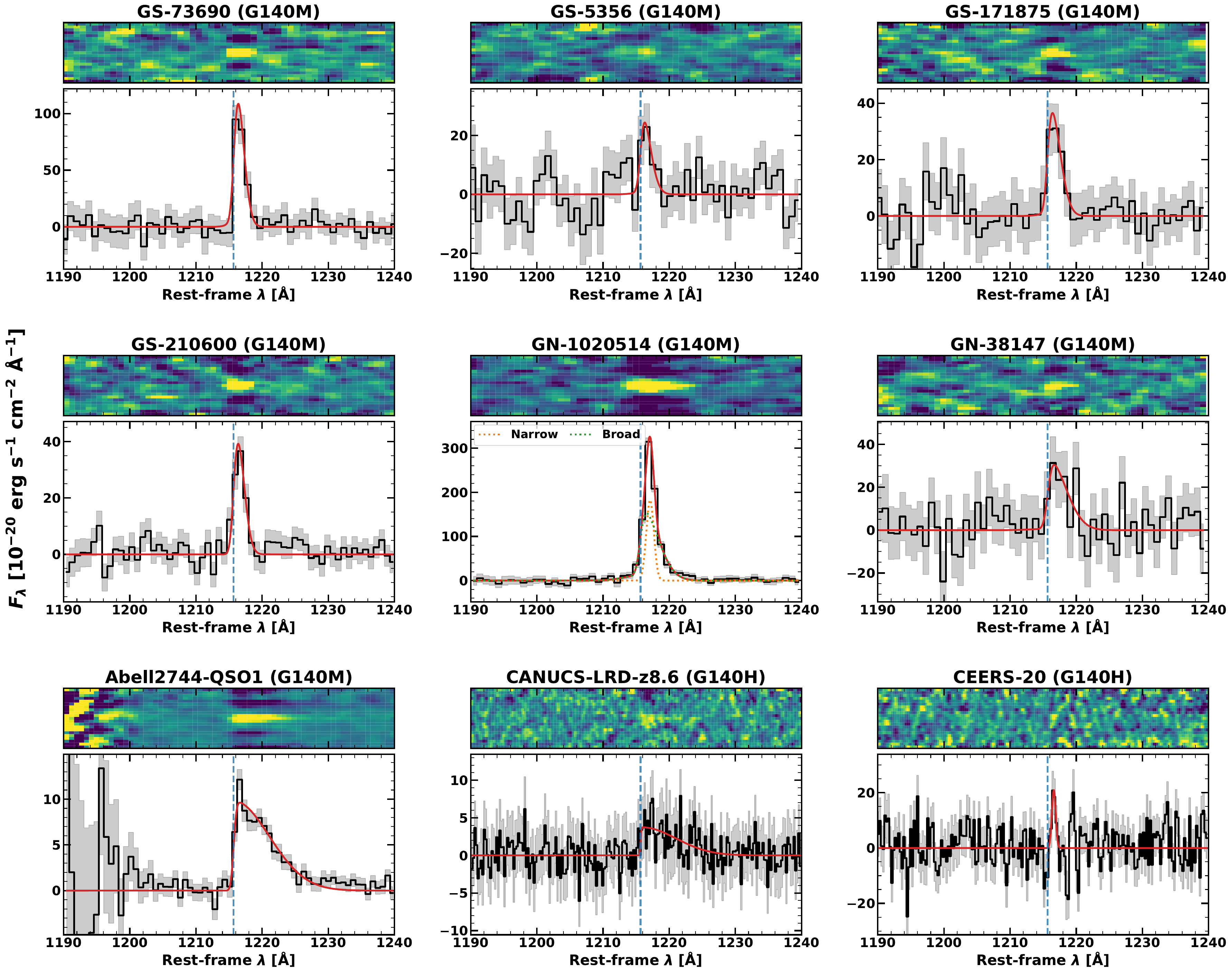}
\caption{NIRSpec Ly$\alpha$ spectra and best-fitting models for the nine LRDs with Ly$\alpha$ detections at S/N$>3$. The two-dimensional spectra are shown above the extracted one-dimensional spectra. For display, the two-dimensional spectra are smoothed along the wavelength direction with a Gaussian kernel of $\sigma=1$ spectral pixel. The black lines and gray shading show the one-dimensional spectra and their $1\sigma$ uncertainties, respectively. The red curves show the best-fitting models, and the dashed lines indicate the Ly$\alpha$ wavelengths at the systemic redshifts. For GN-1020514, the orange and green curves show the narrow and broad components, respectively.}
\label{fig:jwst_lrd_lya_spectra}
\end{figure*}

The fitting results for the MUSE and NIRSpec LRD samples are listed in Table~\ref{tab:lrd_spectroscopic_sample}.
The reported 68\% credible intervals and upper limits on the Ly$\alpha$ fitting results are derived from the MCMC samples using \texttt{GetDist} \citep{lewis25getdist}.
All five Ly$\alpha$-detected MUSE sources show narrow Ly$\alpha$ emission with FWHM values of $200$--$500\ \mathrm{km\ s^{-1}}$.
For the JWST/NIRSpec sample, we detect Ly$\alpha$ at S/N$>3$ in nine of the 23 LRDs.
Figure~\ref{fig:jwst_lrd_lya_spectra} shows the NIRSpec Ly$\alpha$ spectra and best-fitting models for these nine detected sources only.
Abell2744-QSO1 and CANUCS-LRD-z8.6 show broad Ly$\alpha$ emission with $\mathrm{FWHM}_{\mathrm{Ly}\alpha}>1000\ \mathrm{km\ s^{-1}}$, as reported in previous studies \citep{xji26, morishita26, tang26}.
The other seven detected sources show narrow Ly$\alpha$ emission with $\mathrm{FWHM}_{\mathrm{Ly}\alpha}<1000\ \mathrm{km\ s^{-1}}$.
GN-38147 has a relatively broad $\mathrm{FWHM}_{\mathrm{Ly}\alpha}=915^{+200}_{-300}\ \mathrm{km\ s^{-1}}$, but the uncertainty is large.

The WAIC criterion selects the double-Gaussian model only for GN-1020514.
The broad component has $\Delta v<98.1\ \mathrm{km\ s^{-1}}$ and $\mathrm{FWHM}_{\mathrm{Ly}\alpha}=841^{+31}_{-200}\ \mathrm{km\ s^{-1}}$.
Unlike the broad Ly$\alpha$ emission in Abell2744-QSO1 and CANUCS-LRD-z8.6, this component does not have $\mathrm{FWHM}_{\mathrm{Ly}\alpha}>1000\ \mathrm{km\ s^{-1}}$ and does not by itself provide clear evidence for an AGN origin.
GN-1020514 is the brightest LRD in both $M_{\mathrm{UV}}$ and Ly$\alpha$ flux in our sample, and its high S/N may be what reveals the non-Gaussianity of its Ly$\alpha$ profile.

For the three Ly$\alpha$-detected MUSE sources that are also included in the NIRSpec sample, Table~\ref{tab:lrd_spectroscopic_sample} lists the results from both fits.
Their Ly$\alpha$ fluxes do not agree, likely because of NIRSpec slit losses.
For a source with extended UV emission such as GS-204851, components other than the clump hosting the LRD may also contribute significantly to the Ly$\alpha$ flux measured by MUSE.
Ly$\alpha$ is detected in both datasets for GS-210600, whereas GS-9698 and GS-204851 are detected only with MUSE.
The latter difference may reflect the longer integration time of MUSE-Deep as well as slit losses, which arise because the MUSE spectra capture Ly$\alpha$ emission over a spatially broader region than the NIRSpec spectra.
However, if these MUSE LRDs hosted broad Ly$\alpha$ emission similar to that in Abell2744-QSO1 or CANUCS-LRD-z8.6, most of its flux would lie in the broad wings outside the narrow Ly$\alpha$ core.
We therefore expect the different spatial coverage of MUSE and NIRSpec to have a negligible effect on the detectability of such broad Ly$\alpha$ emission.

Finally, most NIRSpec sources are observed with G140M, whereas CANUCS-LRD-z8.6 and CEERS-20 are observed with G140H. The MUSE spectra have a higher spectral resolution than either grating.
For the G140M spectra, we calculate the LSF with \texttt{msafit}, assuming that the LRD is a point source at its measured position in the MSA slit \citep{degraaff24msafit}.
These point-source LSFs have $R\sim1000$, corresponding to a velocity FWHM of $\sim300\ \mathrm{km\ s^{-1}}$.
They differ substantially from the nominal G140M LSF, which assumes uniform slit illumination and has $R\sim600$, corresponding to a velocity FWHM of $\sim500\ \mathrm{km\ s^{-1}}$ at the observed Ly$\alpha$ wavelengths.
The point-source LSF width is comparable to the FWHM values measured for the narrow Ly$\alpha$ lines, so the narrow Ly$\alpha$ FWHM values measured from the G140M spectra carry a systematic uncertainty from the LSF modeling and should be interpreted with caution.
The FWHM measurements are more robust for the higher-resolution MUSE spectra ($R\sim3000$) and for the NIRSpec G140H spectra of CEERS-20 and CANUCS-LRD-z8.6, as well as for the broad Ly$\alpha$ emission in Abell2744-QSO1, whose FWHM is much larger than the G140M LSF width.

\section{Stacked Spectra}
\label{sec:stacked_spectra}

\citet{geris26} reported broad Ly$\alpha$ emission with $\mathrm{FWHM}\sim1200\ \mathrm{km\ s^{-1}}$ in a stack of the JWST/NIRSpec G140M spectra of eight LRDs. Motivated by this result, we use the larger JWST sample constructed in this work and the higher-resolution MUSE sample to investigate a possible hidden AGN contribution to the Ly$\alpha$ emission.

We stack all eight MUSE LRDs listed in Table~\ref{tab:lrd_spectroscopic_sample} and, after excluding Abell2744-QSO1 and CANUCS-LRD-z8.6 with individually detected broad Ly$\alpha$ emission, the remaining 21 NIRSpec LRDs listed in the same table. Following \citet{geris26}, we construct unnormalized median stacks. We shift the spectra to the rest frame using the systemic redshifts and match their spectral resolutions within each sample to that of the lowest-resolution spectrum, $R=1768$ for MUSE and $R=832$ for NIRSpec. We use \texttt{SpectRes} \citep{carnall17} to resample the spectra onto a common rest-frame wavelength grid. For the MUSE stack, we use the wavelength grid of the lowest-resolution spectrum, whereas for the NIRSpec stack we use a grid sampled at three bins per resolution element. We estimate the uncertainty by perturbing each resampled spectrum according to its flux uncertainty and forming median stacks over 1000 realizations for both the MUSE and NIRSpec samples. We adopt the mean and standard deviation of these realizations as the stacked spectrum and its uncertainty. For comparison, we apply the same stacking procedure to the eight NIRSpec LRDs analyzed by \citet{geris26}.

\begin{figure*}[t]
\centering
\includegraphics[width=\textwidth]{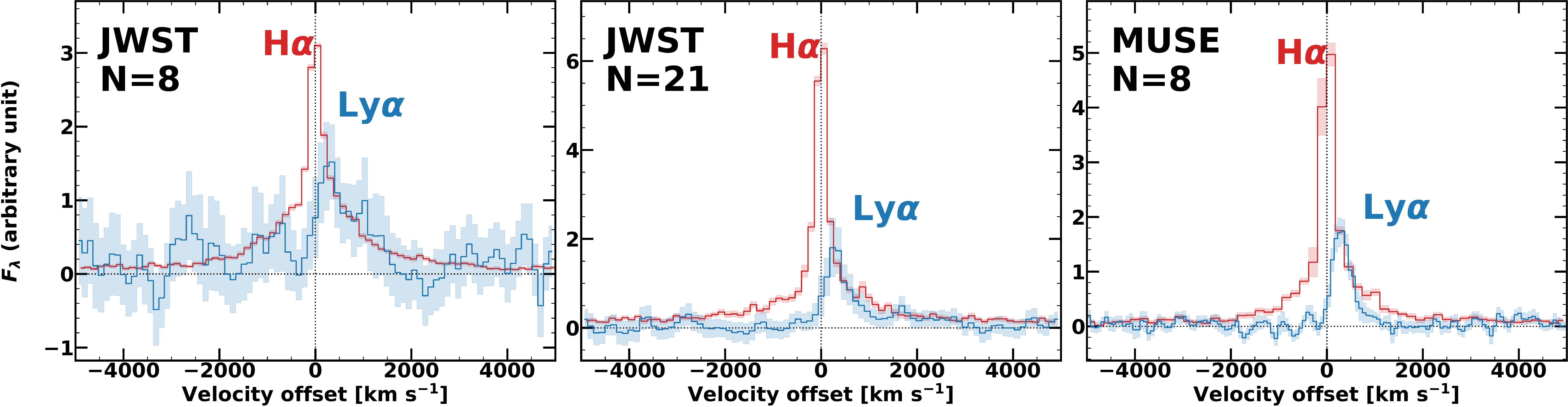}
\caption{Normalized stacked Ly$\alpha$ and H$\alpha$ profiles for the eight-source JWST/NIRSpec LRD sample used by \citet{geris26} (left), the 21-source JWST/NIRSpec LRD sample listed in Table~\ref{tab:lrd_spectroscopic_sample} after excluding Abell2744-QSO1 and CANUCS-LRD-z8.6 (middle), and the eight-source MUSE LRD sample listed in the same table (right). The profiles are normalized at a velocity offset of $500\ \mathrm{km\ s^{-1}}$ for display. The blue and red lines show the Ly$\alpha$ and H$\alpha$ profiles, respectively, and the shaded regions show their $1\sigma$ uncertainties.}
\label{fig:lrd_stack_profiles}
\end{figure*}

The left, middle, and right panels of Figure~\ref{fig:lrd_stack_profiles} show the Ly$\alpha$ and H$\alpha$ profiles for the NIRSpec sample analyzed by \citet{geris26}, the 21-source NIRSpec sample constructed in this work from the sources listed in Table~\ref{tab:lrd_spectroscopic_sample}, and the MUSE sample, respectively. The profiles are normalized by their fluxes at a velocity offset of $500\ \mathrm{km\ s^{-1}}$ for display. For the eight-source sample of \citet{geris26}, the Ly$\alpha$ and H$\alpha$ profiles are similar over $0$--$1500\ \mathrm{km\ s^{-1}}$ despite the large uncertainties, suggesting a relatively broad Ly$\alpha$ profile. In our 21-source JWST/NIRSpec stack with higher S/N, the Ly$\alpha$ flux decreases steadily over $500$--$1500\ \mathrm{km\ s^{-1}}$, and no wing corresponding to the broad H$\alpha$ component is clearly seen. The MUSE stack, which benefits from both the higher spectral resolution and the longer exposure time per source, is more conclusive. Although the stacked Ly$\alpha$ profile has a high S/N, it shows no broad Ly$\alpha$ emission corresponding to the broad H$\alpha$ component. These higher-quality stacked spectra therefore suggest that broad Ly$\alpha$ emission is absent, which we test quantitatively below.

\begin{figure*}[t]
\centering
\includegraphics[width=\textwidth]{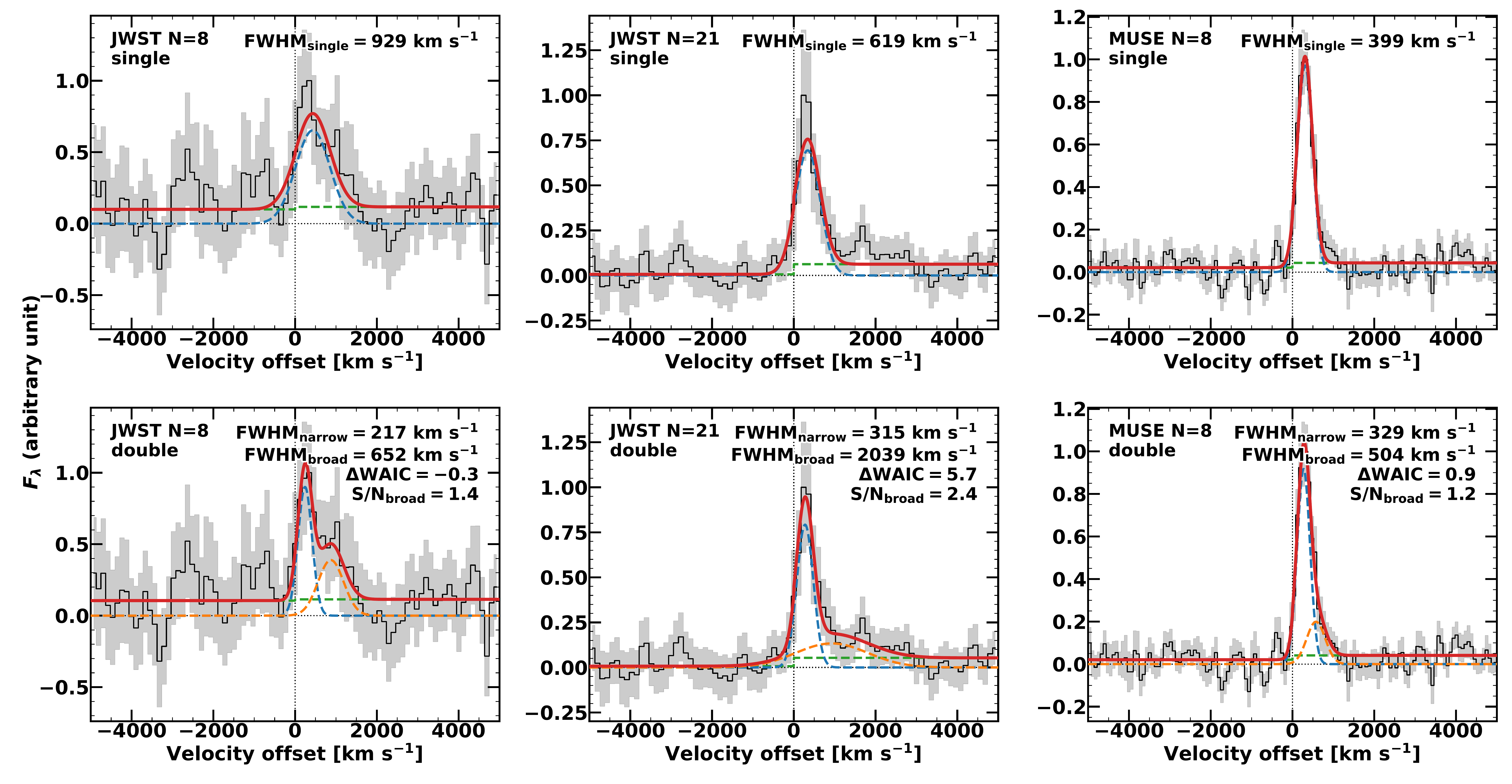}
\caption{Single- and double-Gaussian fits to the stacked Ly$\alpha$ profiles. The left, middle, and right columns show the same eight-source JWST/NIRSpec, 21-source JWST/NIRSpec, and eight-source MUSE samples as in Figure~\ref{fig:lrd_stack_profiles}, respectively. The top and bottom rows show the single- and double-Gaussian fits, respectively. The spectra are normalized by their observed peak fluxes. The black lines and gray shading show the stacked spectra and their $1\sigma$ uncertainties, respectively. The red curves show the total best-fitting models, the green dashed lines show the continua, and the blue and orange dashed curves show the narrow and broad Gaussian components, respectively. Each panel shows the best-fitting FWHM of the corresponding Gaussian component. We define $\Delta\mathrm{WAIC}=\mathrm{WAIC}_{\mathrm{single}}-\mathrm{WAIC}_{\mathrm{double}}$, and $\mathrm{S/N}_{\mathrm{broad}}$ denotes the signal-to-noise ratio of the broad-component flux measured from its posterior. The single-Gaussian model is selected for all three stacks because none of them reaches $\Delta\mathrm{WAIC}>11.8$, and the broad component is undetected in all of them ($\mathrm{S/N}_{\mathrm{broad}}<3$).}
\label{fig:lrd_stack_lya_fits}
\end{figure*}

For each of the three stacked Ly$\alpha$ spectra, we fit the spectral range within $\pm10000\ \mathrm{km\ s^{-1}}$ of the systemic Ly$\alpha$ wavelength with single- and double-Gaussian models, sampling the posterior with \texttt{emcee} \citep{foremanmackey13} following the Ly$\alpha$ fitting procedure described in Section~\ref{sec:lya_fitting}. We include independent flat continua on the blue and red sides of the Ly$\alpha$ wavelength, and we convolve the line profiles with a Gaussian LSF corresponding to the matched spectral resolution of each stack. Because each stack combines sources at different redshifts, we do not include the IGM attenuation term. We again adopt the double-Gaussian model only when the broad component is detected at S/N$>3$ and $\Delta\mathrm{WAIC}=\mathrm{WAIC}_{\mathrm{single}}-\mathrm{WAIC}_{\mathrm{double}}>11.8$.
To compare the two lines on the same basis, we also fit the three stacked H$\alpha$ spectra with the same single- and double-Gaussian models and the same selection criteria.

Figure~\ref{fig:lrd_stack_lya_fits} shows the best-fitting single- and double-Gaussian Ly$\alpha$ models for the three stacks. Neither Ly$\alpha$ selection criterion is met for any of them, so we adopt the single-Gaussian model for Ly$\alpha$, whereas the double-Gaussian model is selected for all three H$\alpha$ stacks. For the sample of \citet{geris26}, we obtain a relatively broad Ly$\alpha$ profile with $\mathrm{FWHM}_{\mathrm{Ly}\alpha}=929^{+543}_{-199}\ \mathrm{km\ s^{-1}}$, which is consistent with the $\mathrm{FWHM}\sim1200\ \mathrm{km\ s^{-1}}$ that they report for the same eight sources, although our measurement has a large uncertainty owing to the lower signal-to-noise ratio of the Ly$\alpha$ flux in this stack ($4.0\sigma$). For the larger JWST/NIRSpec sample and the MUSE sample constructed in this work, we obtain $\mathrm{FWHM}_{\mathrm{Ly}\alpha}=619^{+289}_{-140}$ and $399^{+35}_{-39}\ \mathrm{km\ s^{-1}}$, respectively. These values are much smaller than the FWHM values of the broad H$\alpha$ components of the same two stacks, $1709^{+121}_{-140}$ and $1522^{+124}_{-99}\ \mathrm{km\ s^{-1}}$, and we find no sign of a broad Ly$\alpha$ component associated with broad H$\alpha$. We therefore conclude that the stacked Ly$\alpha$ spectra show no evidence of a hidden AGN contribution.

We caution that Ly$\alpha$ has a different velocity offset in each source, so stacking multiple spectra broadens the line profile. The FWHM values of $\sim400$--$620\ \mathrm{km\ s^{-1}}$ measured for the narrow Ly$\alpha$ emission in the larger JWST/NIRSpec and MUSE stacks therefore do not represent the typical narrow Ly$\alpha$ FWHM of individual LRDs. The source-to-source variation in the Ly$\alpha$ velocity offset can also smear out a possible broad Ly$\alpha$ component and thereby limit its detectability. This is an inherent limitation of stacking Ly$\alpha$ spectra at the systemic redshifts. We therefore focus on individual sources below.

\section{Results}
\label{sec:results}

\subsection{Broad Ly\texorpdfstring{$\alpha$}{alpha} Fraction}
\label{sec:broad_lya_fraction}

As described in Section~\ref{sec:lya_fitting}, we identify no new broad Ly$\alpha$ emitters beyond the previously reported Abell2744-QSO1 and CANUCS-LRD-z8.6.
This result suggests that LRDs in which Ly$\alpha$ photons escape from the broad-line region represent a rare subset of the overall LRD population.
A naive estimate based on the two broad Ly$\alpha$ emitters among the 27 distinct LRDs in Table~\ref{tab:lrd_spectroscopic_sample} gives a fraction of $2/27\simeq7\%$.
However, this estimate does not account for variations in detection sensitivity.
In particular, the broad Ly$\alpha$ emission of Abell2744-QSO1 is exceptionally faint and is detected because of its large lensing magnification and the ultra-deep SPURS observations.
Broad Ly$\alpha$ emission of the same luminosity would be difficult to detect in most of the other LRDs.

To account for the different depths of the individual spectra, we determine a broad Ly$\alpha$ detection limit for each source before measuring the broad Ly$\alpha$ emitter fraction for our LRD sample, which spans $-20.2<M_{\mathrm{UV}}<-16.8$, and for the comparison SFG samples.
For sources without a detected broad component, we estimate the sensitivity by fixing the line width to that of CANUCS-LRD-z8.6, $\mathrm{FWHM}_{\mathrm{Ly}\alpha}=1922\ \mathrm{km\ s^{-1}}$.
If $N$ spectral bins fall within $\pm\mathrm{FWHM}$ of the expected Ly$\alpha$ wavelength, the $1\sigma$ line-flux limit is calculated as $F_{1\sigma}=f\,\Delta\lambda_{\mathrm{bin}}\sqrt{N}$, where $f$ is the flux-density uncertainty per wavelength bin and $\Delta\lambda_{\mathrm{bin}}$ is the bin width.
We adopt three times this value as the $3\sigma$ upper limit.
This approximation agrees with the MCMC flux uncertainties to within 2\% for Abell2744-QSO1 and 10\% for CANUCS-LRD-z8.6.
In Section~\ref{sec:lya_fitting}, when fitting Ly$\alpha$ with narrow and broad components, we require not only that the broad component has S/N$>3$ but also that the double-Gaussian model satisfies $\Delta\mathrm{WAIC}>11.8$.
The detection limits here are based only on the $3\sigma$ flux upper limits.
To assess the effect of the additional WAIC criterion, we fit mock spectra that include narrow and broad Ly$\alpha$ components and noise.
We find that the S/N$=3$ and $\Delta\mathrm{WAIC}=11.8$ thresholds nearly coincide, indicating that the additional WAIC criterion has little effect on the estimated detection limits.
For the JADES SFGs described in Section~\ref{sec:sfg_sample}, we fit Ly$\alpha$ using the same method as in Section~\ref{sec:lya_fitting}.
For the MUSE SFGs, we examine the Ly$\alpha$ FWHM values in the MUSE-Deep emission-line catalog.
Neither SFG sample contains a broad Ly$\alpha$ emitter, so all of the SFG broad-component measurements are upper limits.
We apply the same sensitivity calculation to the MUSE and JWST LRDs and to the MUSE and JADES SFG samples.
After removing sources duplicated between the MUSE and JWST samples, the calculation includes 27 LRDs and 279 SFGs.

\begin{figure*}[t]
\centering
\includegraphics[width=\textwidth]{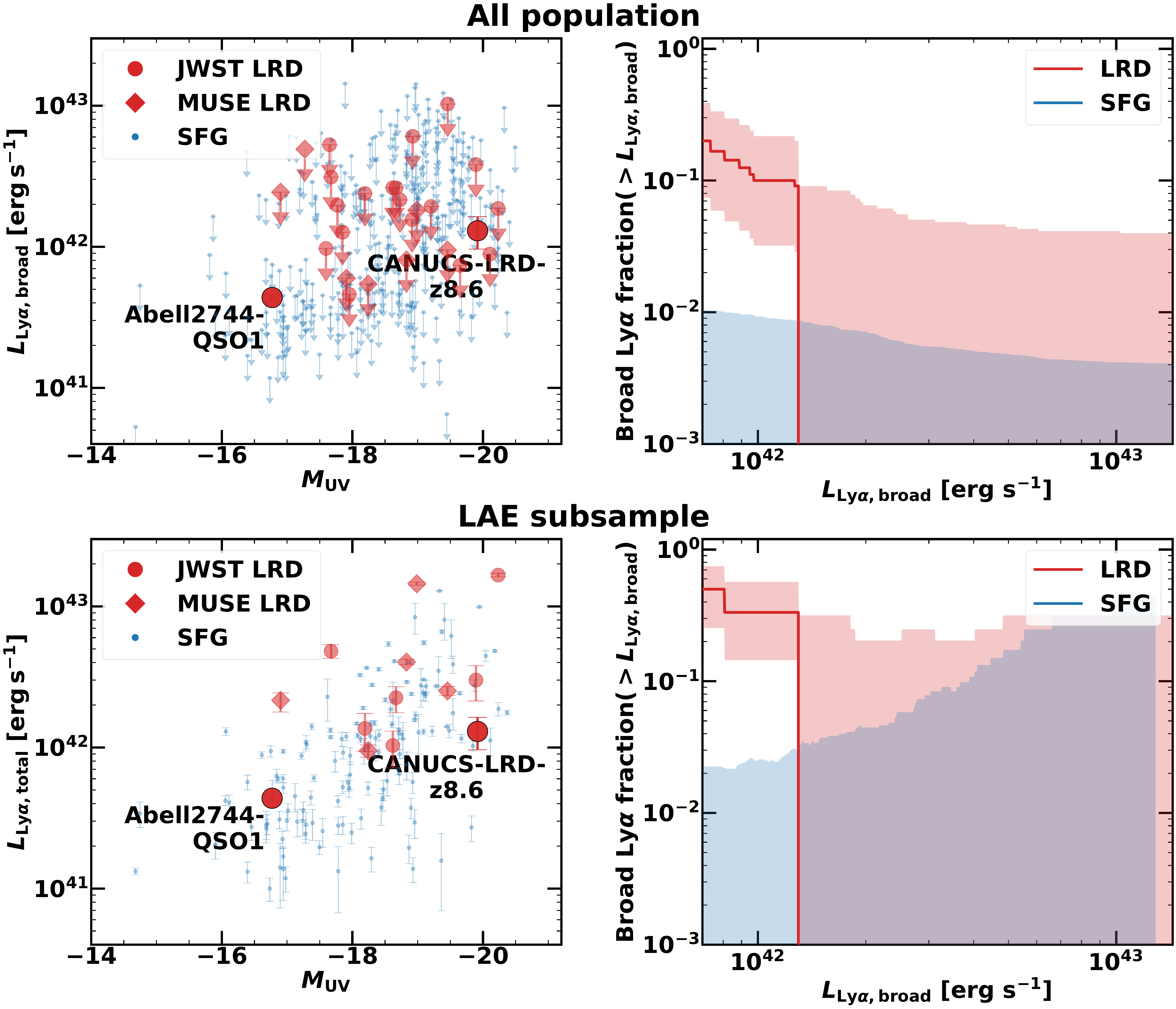}
\caption{Broad Ly$\alpha$ emitter fractions in the whole population (top row) and the Ly$\alpha$-emitter (LAE) subsample (bottom row). The left panels show Ly$\alpha$ luminosity versus rest-frame UV absolute magnitude. The top-left panel shows the broad Ly$\alpha$ luminosity. CANUCS-LRD-z8.6 and Abell2744-QSO1 are the two secure broad Ly$\alpha$ detections, while all other sources are shown as upper limits. The bottom-left panel shows the total Ly$\alpha$ luminosity, including both the narrow and broad components. The red circles, red diamonds, and blue circles represent JWST LRDs, MUSE LRDs, and SFGs, respectively. The right panels show the broad Ly$\alpha$ emitter fraction $f_{\mathrm{broad}}$ versus the minimum broad Ly$\alpha$ luminosity $L$. The lines and shaded regions show the inferred values and their 68\% intervals, respectively. At $L=10^{42}\ \mathrm{erg\ s^{-1}}$, $f_{\mathrm{broad}}$ is $0.10_{-0.07}^{+0.12}$ for the LRDs and $0.02$ (95\% upper limit) for the SFGs in the whole population, and $0.33_{-0.19}^{+0.24}$ for the LRDs and $0.06$ (95\% upper limit) for the SFGs in the LAE subsample. In both cases, $f_{\mathrm{broad}}$ of the LRDs is higher than that of the SFGs.}
\label{fig:broad_lya_fraction_summary}
\end{figure*}

We define the broad Ly$\alpha$ emitter fraction $f_{\mathrm{broad}}$ at luminosity $L$ as the fraction of sources with broad Ly$\alpha$ luminosity at or above $L$.
We calculate this fraction while varying $L$.
At each value of $L$, the denominator $n$ includes non-detections whose $3\sigma$ upper limits satisfy $L_{3\sigma}\leq L$, together with the two secure broad Ly$\alpha$ detections.
The numerator $k$ is the number of detected sources with $L_{\mathrm{Ly}\alpha,\mathrm{broad}}\geq L$.
The posterior of $f_{\mathrm{broad}}$ is determined from the binomial likelihood.
With a uniform prior, the posterior is $\mathrm{Beta}(k+1,n-k+1)$, and for $k>0$, we report its mode and 68\% highest-posterior-density interval (HPDI).
For $k=0$, the posterior mode is zero, and we quote the 95\% one-sided upper limit.
The top row of Figure~\ref{fig:broad_lya_fraction_summary} shows the broad Ly$\alpha$ luminosities and upper limits in the left panel and $f_{\mathrm{broad}}$ in the right panel.
At $L=10^{42}\ \mathrm{erg\ s^{-1}}$, the LRD sample has $k=1$ and $n=10$, giving $0.10_{-0.07}^{+0.12}$, whereas the SFG sample has $k=0$ and $n=121$, giving a 95\% upper limit of $0.02$.
We use the values at $L=10^{42}\ \mathrm{erg\ s^{-1}}$ as the representative $f_{\mathrm{broad}}$ hereafter.
As shown in Figure~\ref{fig:broad_lya_fraction_summary}, $f_{\mathrm{broad}}$ among the LRDs remains higher than that among the SFGs at $L<1.3\times10^{42}\ \mathrm{erg\ s^{-1}}$.
At larger values of $L$, neither LRD has $L_{\mathrm{Ly}\alpha,\mathrm{broad}}\geq L$, so only an upper limit can be placed on $f_{\mathrm{broad}}$ among the LRDs.

We also calculate $f_{\mathrm{broad}}$ in subsamples restricted to Ly$\alpha$ emitters (LAEs) rather than the full LRD and SFG populations.
This restriction reduces the possible effects of attenuation by the interstellar and intergalactic media, which affects both broad and narrow Ly$\alpha$ emission.
We identify nine JWST LRDs, five MUSE LRDs, 124 MUSE SFGs, and 20 JADES SFGs as LAEs.
After removing sources duplicated between the MUSE and JWST samples, the LAE subsamples contain 13 LRDs and 142 SFGs.
For these subsamples, the denominator $n$ contains sources with measured total Ly$\alpha$ luminosity $L_{\mathrm{Ly}\alpha,\mathrm{total}}\geq L$ and a broad-component limit $L_{3\sigma}\leq L$.
The numerator $k$ is again the number with $L_{\mathrm{Ly}\alpha,\mathrm{broad}}\geq L$.
At $L=10^{42}\ \mathrm{erg\ s^{-1}}$, $f_{\mathrm{broad}}$ is $0.33_{-0.19}^{+0.24}$ for the LRD LAEs ($k=1$, $n=3$), compared with a 95\% upper limit of $0.06$ for the SFG LAEs ($k=0$, $n=44$).
The bottom row of Figure~\ref{fig:broad_lya_fraction_summary} shows the total Ly$\alpha$ luminosities used to select the LAE subsamples in the left panel and their $f_{\mathrm{broad}}$ in the right panel.
The value of $f_{\mathrm{broad}}$ among the LRD LAEs is higher than that in the full LRD sample, while that among the SFG LAEs remains consistent with zero.
However, the LRD LAE result is based on only three sources and has a correspondingly wide HPDI.

\subsection{Ly\texorpdfstring{$\alpha$}{alpha} FWHM Distribution}
\label{sec:lya_fwhm_distribution}

\begin{figure}[t]
\centering
\includegraphics[width=\columnwidth]{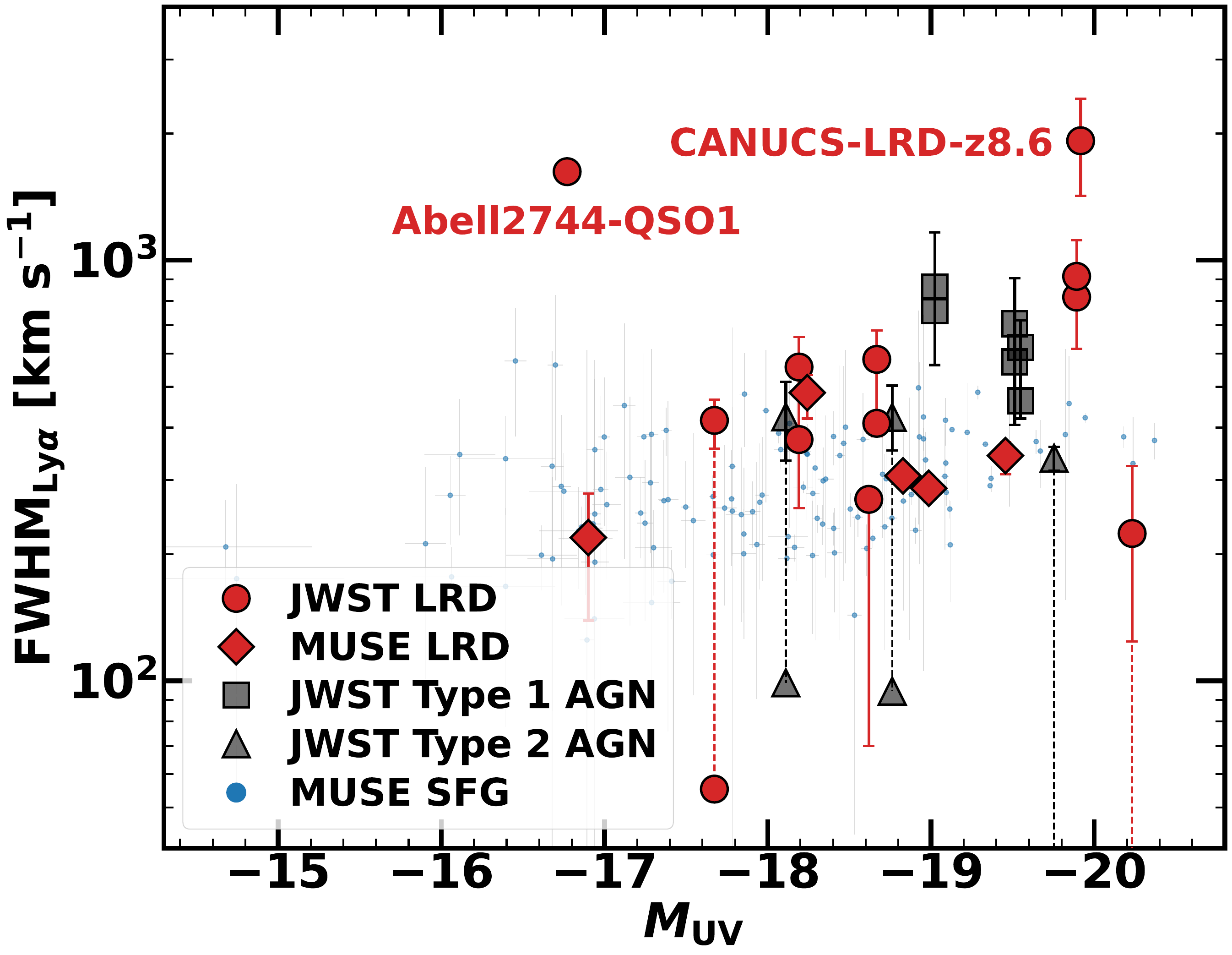}
\caption{Ly$\alpha$ FWHM versus rest-frame UV absolute magnitude. The red circles, red diamonds, black squares, black triangles, and blue circles represent JWST LRDs, MUSE LRDs, JWST Type 1 AGNs, JWST Type 2 AGNs, and MUSE SFGs, respectively. For the JWST sources observed with G140M, except Abell2744-QSO1, whose Ly$\alpha$ FWHM is much larger than the instrumental resolution, the measured FWHM depends on the assumed LSF. For each of these sources, the upper symbol shows the measurement obtained with the point-source LSF calculated using \texttt{msafit}, while the lower symbol shows the measurement obtained with the nominal JWST LSF, and the two measurements are connected by a dashed line. The true FWHM is expected to lie between these two estimates. CANUCS-LRD-z8.6 and CEERS-20 are observed with G140H and are shown with a single symbol.}
\label{fig:muv_lya_fwhm}
\end{figure}

\begin{figure}[t]
\centering
\includegraphics[width=\columnwidth]{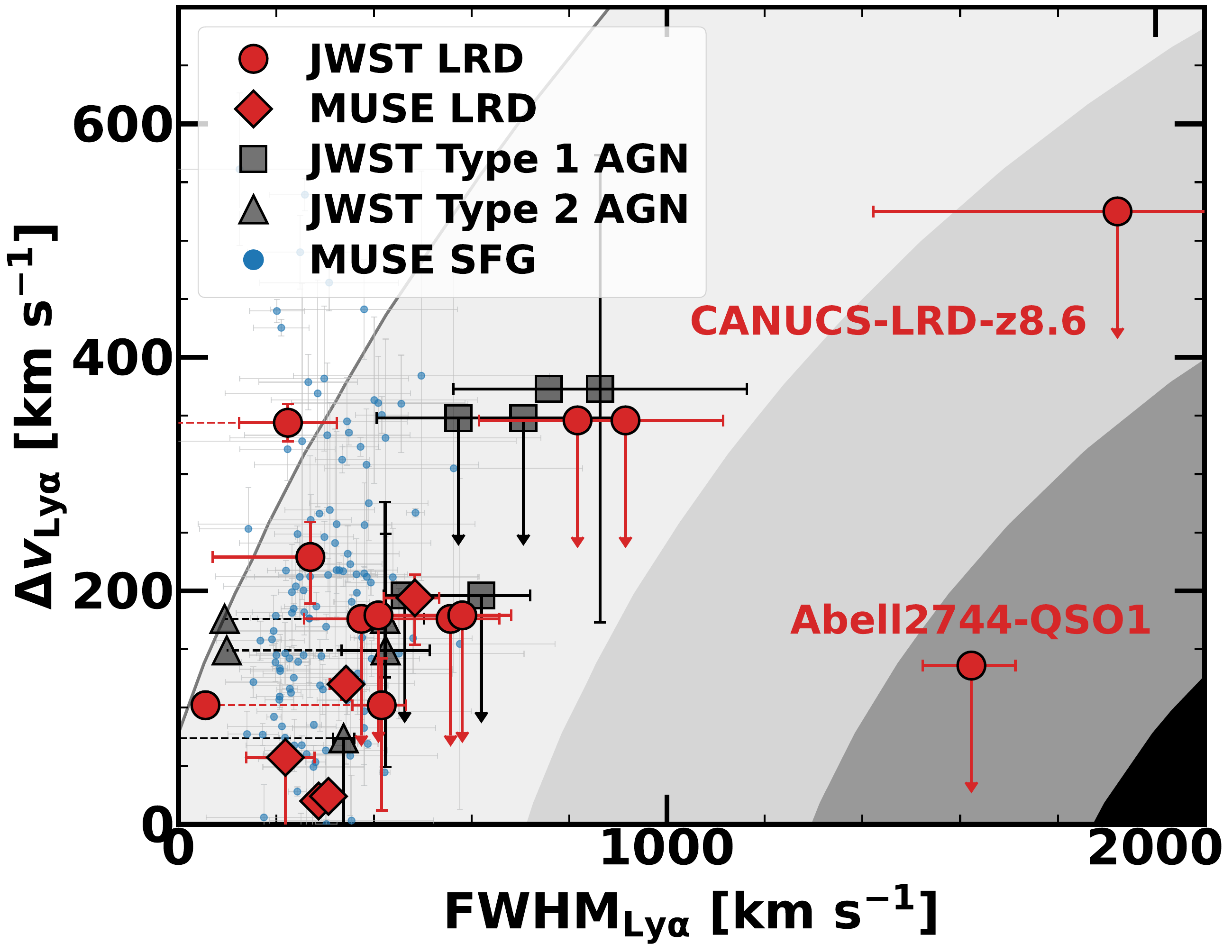}
\caption{Ly$\alpha$ velocity offset versus Ly$\alpha$ FWHM. The red circles, red diamonds, black squares, black triangles, and blue circles represent JWST LRDs, MUSE LRDs, JWST Type 1 AGNs, JWST Type 2 AGNs, and MUSE SFGs, respectively. For the JWST sources observed with G140M, except Abell2744-QSO1, the two FWHM estimates obtained with the \texttt{msafit} and nominal JWST LSFs are connected by a dashed line, as described in Figure~\ref{fig:muv_lya_fwhm}. The gray contours show SDSS DR16Q quasars from \citet{wu22}. From darkest to lightest, the contours correspond to the 20\%, 40\%, 60\%, and 80\% levels of the SDSS QSO distribution. The downward arrows indicate 68\% upper limits on the Ly$\alpha$ velocity offset.}
\label{fig:fwhm_delta_v_lya}
\end{figure}

We now focus on the distribution of the LRD Ly$\alpha$ FWHM measurements presented in Section~\ref{sec:lya_fitting}.
Figures~\ref{fig:muv_lya_fwhm} and \ref{fig:fwhm_delta_v_lya} show $M_{\mathrm{UV}}$ versus Ly$\alpha$ FWHM and Ly$\alpha$ FWHM versus velocity offset, respectively.
Alongside the MUSE and JWST LRDs, the figures include JWST Type 1 and Type 2 AGNs as well as the MUSE-Deep SFGs described in Section~\ref{sec:sfg_sample}.
For these SFGs, we adopt the Ly$\alpha$ FWHM values from the MUSE-Deep emission-line catalog.
We calculate the velocity offset using the Ly$\alpha$ peak wavelength in that catalog and the systemic redshift from the matched DJA NIRSpec catalog.

The two LRDs with broad Ly$\alpha$ emission lie clearly above the SFG distribution shown by the blue points in Figures~\ref{fig:muv_lya_fwhm} and \ref{fig:fwhm_delta_v_lya}, with Ly$\alpha$ lines broader than those of any SFG in the comparison sample.
Figure~\ref{fig:fwhm_delta_v_lya} also shows contours of the Ly$\alpha$ FWHM and velocity-offset distribution of SDSS DR16Q quasars from \citet{wu22}.
Although the two broad Ly$\alpha$ LRDs are inconsistent with the MUSE SFG locus, they occupy the same region as the SDSS quasars.
Their velocity offsets are small relative to their line widths, suggesting that the broad Ly$\alpha$ emission is intrinsically broad and originates in the broad-line region, rather than being produced primarily by strong resonant scattering.

Apart from these two broad Ly$\alpha$ emitters, the MUSE LRDs occupy the same region as the MUSE SFGs, supporting a star-formation origin for their narrow Ly$\alpha$ emission.
The JWST LRDs and the Type 1 and Type 2 AGNs have Ly$\alpha$ FWHM values similar to or somewhat larger than those of the MUSE SFGs.
However, as discussed in Section~\ref{sec:lya_fitting}, the spectral resolution of the NIRSpec G140M data is low enough that the inferred narrow-line FWHM depends sensitively on the adopted LSF.
For sources connected by dashed lines, the upper symbols in Figure~\ref{fig:muv_lya_fwhm} and the right symbols in Figure~\ref{fig:fwhm_delta_v_lya} show our fiducial measurements obtained with the point-source LSF calculated using \texttt{msafit}, while the lower symbols in Figure~\ref{fig:muv_lya_fwhm} and the left symbols in Figure~\ref{fig:fwhm_delta_v_lya} show the values obtained using the nominal JWST LSF.
The substantial shifts between these measurements provide an estimate of the possible systematic uncertainty due to LSF modeling.
Accounting for this uncertainty, the Ly$\alpha$ FWHM values of the JWST LRDs and the Type 1 and Type 2 AGNs remain consistent with the SFG distribution.

\subsection{Ly\texorpdfstring{$\alpha$}{alpha} Equivalent Width}
\label{sec:lya_equivalent_width}

\begin{figure}[t]
\centering
\includegraphics[width=\columnwidth]{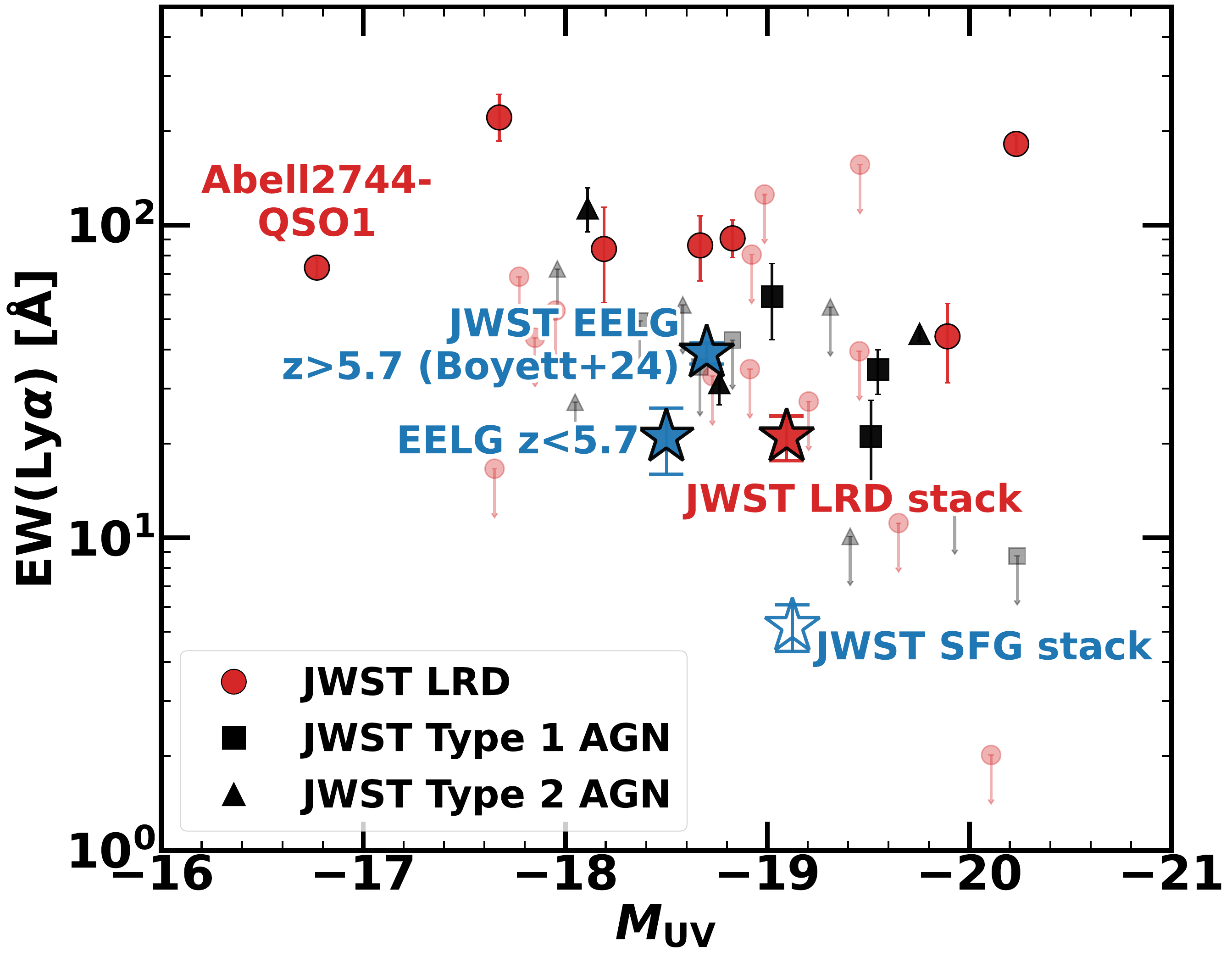}
\caption{Rest-frame Ly$\alpha$ equivalent width versus rest-frame UV absolute magnitude. The red circles, black squares, and black triangles represent JWST LRDs, JWST Type 1 AGNs, and JWST Type 2 AGNs, respectively. The large red star, the two filled blue stars, and the open blue star represent the JWST LRD stack, the $z>5.7$ and $z<5.7$ JADES EELG stacks from \citet{boyett24}, and the JWST SFG stack, respectively. The downward arrows indicate 95\% upper limits. CANUCS-LRD-z8.6 is not included because its continuum is not detected at S/N$>3$ in the G140H spectrum.}
\label{fig:muv_lya_ew}
\end{figure}

We compare the rest-frame Ly$\alpha$ equivalent widths of the JWST LRD and SFG stacks.
The LRD stacks in Section~\ref{sec:stacked_spectra} are unnormalized, following \citet{geris26} to enable a direct comparison with their result.
Here, because the Ly$\alpha$ equivalent width is measured relative to the UV continuum, we instead normalize each LRD and SFG spectrum by its rest-frame 1500~\AA\ continuum before stacking.
This places all spectra on the same rest-UV continuum scale and enables a direct comparison of the relative Ly$\alpha$ strengths of the two populations.

To determine the rest-UV continuum of each LRD in Table~\ref{tab:lrd_spectroscopic_sample}, we fit a power law to its NIRSpec prism spectrum.
We use the rest-frame wavelength range from 1350~\AA\ to the Balmer break at 3646~\AA\ to avoid Ly$\alpha$ damping-wing absorption.
We mask 1440--1590, 1620--1680, and 1860--1980~\AA\ to avoid strong UV emission lines, including N~\textsc{iv}], C~\textsc{iv}, He~\textsc{ii}, and C~\textsc{iii}] \citep{umeda26}.
When a prism spectrum was not obtained at the same slit position as the G140M or G140H spectrum used to measure Ly$\alpha$, we instead fit the corresponding G140M or G140H spectrum.
We extrapolate the best-fitting power law to 1216~\AA\ to estimate the continuum at Ly$\alpha$.

The rest-UV continuum is not detected at S/N$>3$ in the G140M or G140H spectra of CEERS-20, DREAMS-60007, and CANUCS-LRD-z8.6, so we exclude these sources.
We also exclude Abell2744-QSO1, which has individually detected broad Ly$\alpha$ emission.
Applying $M_{\mathrm{UV}}<-18$ to the remaining sources leaves 14 LRDs, whose spectra we normalize by the fitted rest-frame 1500~\AA\ continuum and stack.
Applying the same UV magnitude cut to the SFG sample selects 127 of the 161 JADES SFGs described in Section~\ref{sec:sfg_sample}. We normalize and stack their spectra in the same way.
We then fit Ly$\alpha$ in both normalized stacks following the procedure described in Section~\ref{sec:stacked_spectra}.
Figure~\ref{fig:muv_lya_ew} shows the resulting rest-frame Ly$\alpha$ equivalent widths of the two stacks, as well as the measurements of the individual LRDs and non-LRD Type 1 and Type 2 AGNs, with each stack plotted at the median $M_{\mathrm{UV}}$ of its contributing sources.
The MUSE LRDs are not shown because their MUSE spectra do not cover the rest-frame UV continuum.

The LRD and SFG stacks have similar median UV magnitudes, $M_{\mathrm{UV}}=-18.99$ and $-19.12$, but the Ly$\alpha$ equivalent width of the LRD stack is approximately four times higher than that of the SFG stack.
This Ly$\alpha$ excess is qualitatively consistent with the prism-stack analysis of \citet{geris26}, who also found that the Ly$\alpha$ equivalent width of LRDs is approximately four times that of SFGs at similar UV luminosity.
For comparison, \citet{boyett24} stacked, separately at $z>5.7$ and $z<5.7$, JADES DR1 prism spectra of extreme emission-line galaxies (EELGs) selected to have rest-frame [O~\textsc{iii}] $\lambda5007$ equivalent width greater than $750$~\AA.
The two EELG stacks lie close to the JWST LRD stack in Figure~\ref{fig:muv_lya_ew}.
This similarity suggests that the narrow Ly$\alpha$ emission in LRDs can be produced by vigorous star formation in their host galaxies, accompanied by interstellar-medium conditions that facilitate the escape of Ly$\alpha$ photons, as in EELGs.
Nevertheless, photoionization by the central engine in an AGN narrow-line region cannot be ruled out.

\section{Discussion}
\label{sec:discussion}

In Section~\ref{sec:broad_lya_fraction}, we found that broad Ly$\alpha$ emission is present in only a minority of LRDs, with a representative fraction of approximately 10\%. The only sources with detected broad Ly$\alpha$ emission in Table~\ref{tab:lrd_spectroscopic_sample} are Abell2744-QSO1 and CANUCS-LRD-z8.6. The high redshift of CANUCS-LRD-z8.6 and the lack of a rest-frame optical grating spectrum make precise measurements of its optical continuum and broad Balmer lines difficult. We therefore focus on Abell2744-QSO1 and examine its broad H$\alpha$ luminosity relative to its bolometric luminosity. For comparison, we also derive the broad H$\alpha$ and bolometric luminosities of the other Ly$\alpha$-emitting LRDs in our sample.

\begin{figure}[t]
\centering
\includegraphics[width=\columnwidth]{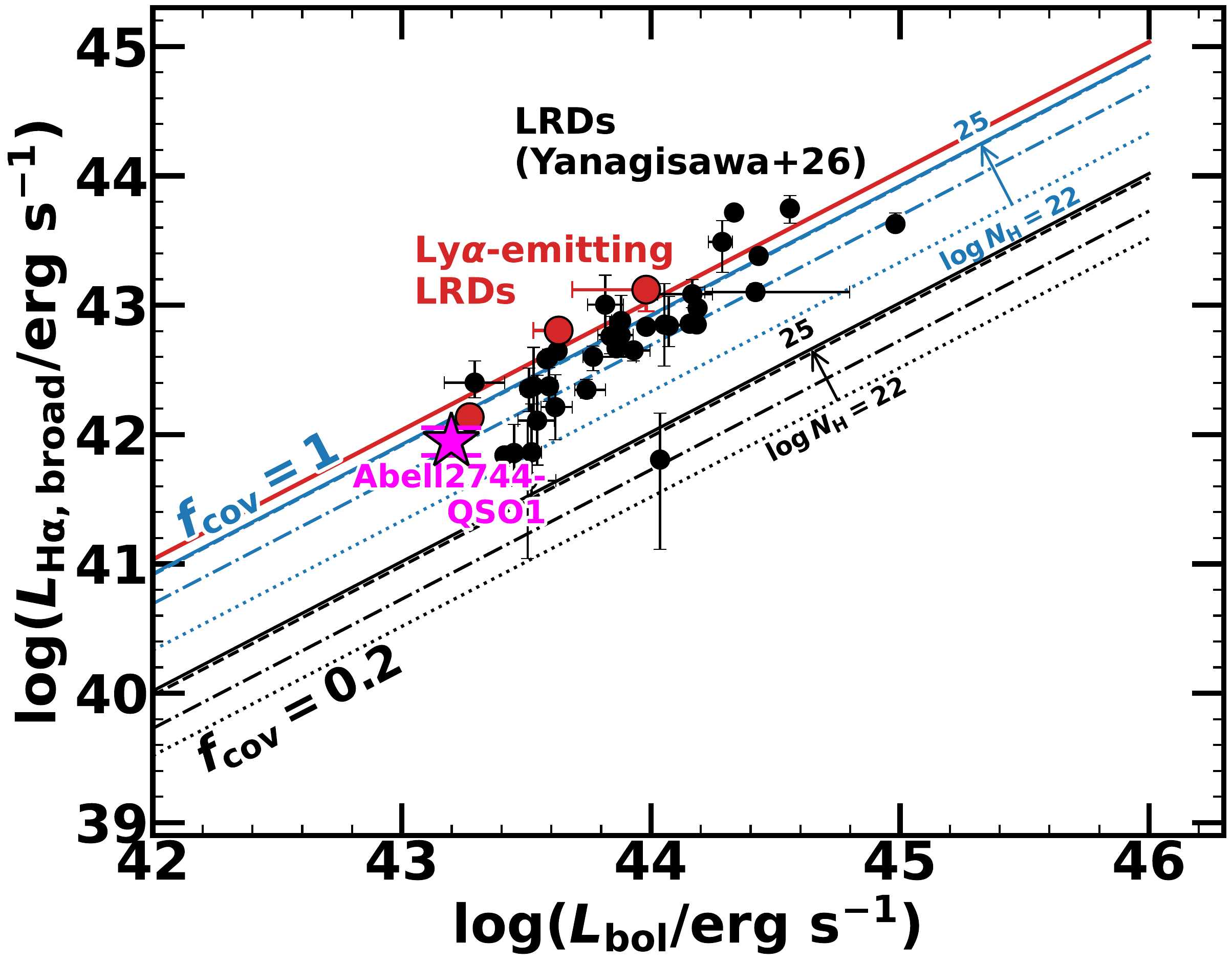}
\caption{Broad H$\alpha$ luminosity versus bolometric luminosity. The black circles show the LRD measurements from \citet{yanagisawa26}. The magenta star and the red circles show the Ly$\alpha$-emitting LRDs of this work, namely Abell2744-QSO1 and GS-13329, GS-204851, and GN-38147, respectively. The red line is the best fit to the three Ly$\alpha$-emitting LRDs other than Abell2744-QSO1. The blue and black lines show the Cloudy predictions from \citet{yanagisawa26} for covering factors of $f_{\mathrm{cov}}=1$ and 0.2, respectively. The line styles indicate hydrogen column densities from $\log(N_{\mathrm{H}}/\mathrm{cm}^{-2})=22$ (dotted) to 25 (solid).}
\label{fig:cloudy_lbol_lha}
\end{figure}

We adopt the broad H$\alpha$ luminosity measured from the SPURS spectrum by \citet{tang26}. To estimate the bolometric luminosity, we assume that the continuum spectrum of Abell2744-QSO1 is the sum of a host-galaxy power law and a modified blackbody associated with the central engine, following \citet{yanagisawa26}. We identify the luminosity of the modified-blackbody component with the bolometric luminosity. We apply the same continuum decomposition to the other Ly$\alpha$-emitting LRDs and take their broad H$\alpha$ luminosities from the fits described in Section~\ref{sec:systemic_redshift}. Because our LRD sample is selected from NIRCam photometry (Section~\ref{sec:lrd_photometric_selection}) rather than from the spectroscopy available for each source, the data required for these two measurements are not in hand for every Ly$\alpha$-emitting LRD. Some sources have no grating spectrum covering H$\alpha$ at their systemic redshift, and for others the prism spectrum does not reach the signal-to-noise ratio in the rest-frame optical continuum that is needed to constrain the modified-blackbody component. We are therefore able to measure both luminosities for three sources in addition to Abell2744-QSO1, namely GS-13329, GS-204851, and GN-38147. Figure~\ref{fig:cloudy_lbol_lha} compares these four Ly$\alpha$-emitting LRDs with the LRD measurements of \citet{yanagisawa26} and their Cloudy predictions for the relation between bolometric and broad H$\alpha$ luminosities as the BLR gas covering factor, $f_{\mathrm{cov}}$, and hydrogen column density are varied.

As found for the LRDs of \citet{yanagisawa26}, the broad H$\alpha$ emission of all four Ly$\alpha$-emitting LRDs is too luminous relative to their bolometric luminosities to be explained by $f_{\mathrm{cov}}=0.2$, the value typical of local Type 1 AGNs, which indicates that the gas enveloping the central engine has a high covering factor in these sources as well. Balmer breaks of non-stellar origin and Balmer-line absorption are commonly observed in LRDs \citep{naidu25, juodzbalis26b, matthee26, yanagisawa26b}, which indicates that the column density of this gas is also high. For Abell2744-QSO1 in particular, \citet{xji25} used the strength of the Balmer break to infer $N_{\mathrm{H}}>10^{23}\ \mathrm{cm}^{-2}$. Among the four Ly$\alpha$-emitting LRDs, however, Abell2744-QSO1 has the smallest ratio of broad H$\alpha$ to bolometric luminosity, $L_{\mathrm{H}\alpha,\mathrm{broad}}/L_{\mathrm{bol}}=0.056$, compared with 0.073 for GS-13329, 0.14 for GN-38147, and 0.15 for GS-204851. Fitting the three sources other than Abell2744-QSO1 with the slope fixed to that of the Cloudy predictions gives $L_{\mathrm{H}\alpha,\mathrm{broad}}/L_{\mathrm{bol}}=0.11$, shown by the red line in Figure~\ref{fig:cloudy_lbol_lha}, which is about twice the value of Abell2744-QSO1. Because this ratio increases with both the covering factor and the hydrogen column density in the Cloudy calculations, the comparatively small ratio of Abell2744-QSO1 may indicate that its covering factor or its column density is somewhat lower than in the other LRDs, leaving an escape path through which broad Ly$\alpha$ photons can reach the observer. This is consistent with Abell2744-QSO1 being the only one of the four in which broad Ly$\alpha$ emission is detected, whereas the Ly$\alpha$ emission of the other three sources is narrow.

In addition, the covering factor of Abell2744-QSO1 obtained by fitting its Balmer absorption lines is $\sim0.27$--$0.56$ \citep{juodzbalis26b, deugenio26abs}, the lowest value among the JWST LRDs with Balmer absorption measurements \citep{yanagisawa26b}. The covering factor of the $n=2$ hydrogen gas derived from the absorption lines and the covering factor of the BLR clouds discussed here are different quantities, so the two cannot be compared rigorously. This nonetheless supports the interpretation that the covering factor of Abell2744-QSO1 is comparatively low. Interpolating the Cloudy calculations of \citet{yanagisawa26} gives $f_{\mathrm{cov}}=0.71^{+0.17}_{-0.13}$ when $\log(N_{\mathrm{H}}/\mathrm{cm}^{-2})$ is fixed to 24, or $\log(N_{\mathrm{H}}/\mathrm{cm}^{-2})=23.18^{+0.41}_{-0.39}$ when $f_{\mathrm{cov}}$ is fixed to unity. We note that these estimates should not be interpreted as precise constraints because $L_{\mathrm{H}\alpha,\mathrm{broad}}/L_{\mathrm{bol}}$ also depends on the gas volume density, ionization parameter, and other model parameters. The bolometric luminosity has an additional uncertainty because Abell2744-QSO1 lies at $z\sim7$, where the NIRSpec/prism spectrum alone does not constrain the temperature of the modified-blackbody component precisely.

\begin{figure*}[t]
\centering
\includegraphics[width=\textwidth]{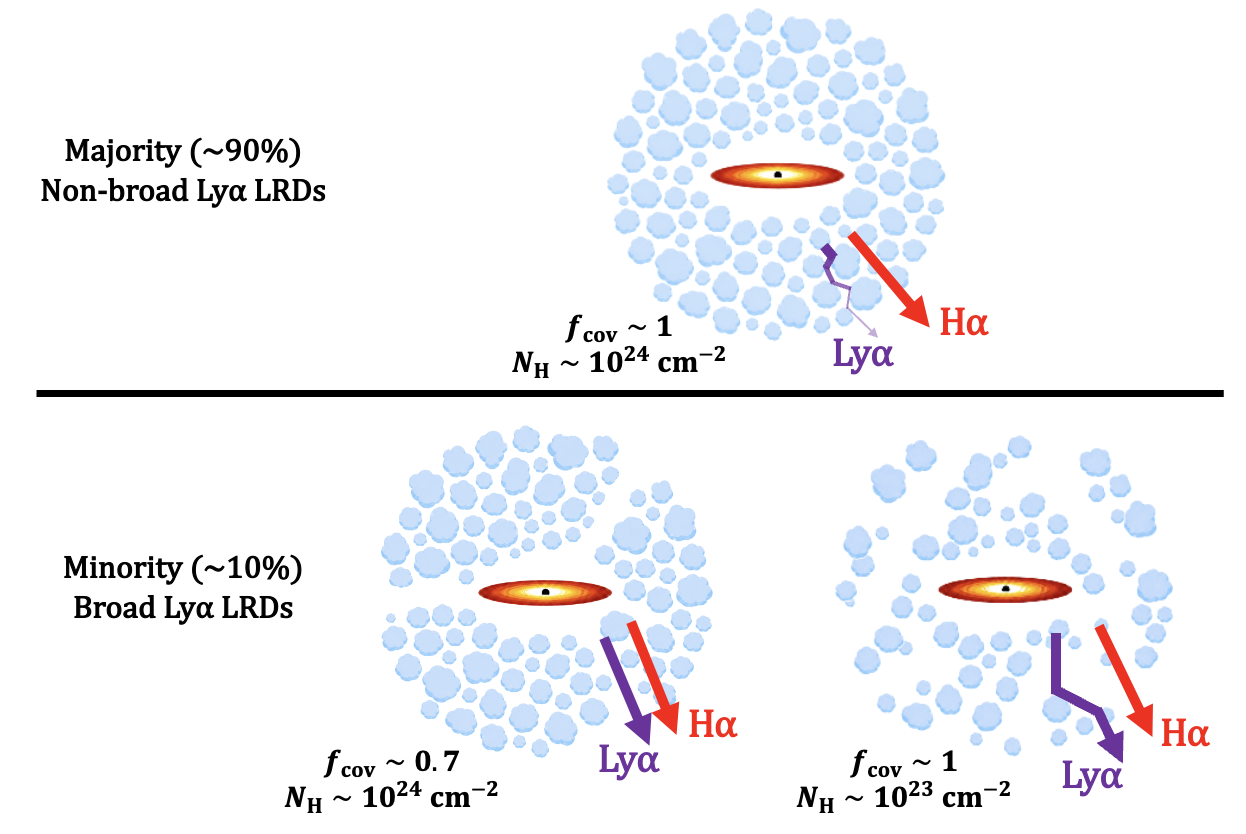}
\caption{Schematic interpretation of the visibility of broad Ly$\alpha$ emission in LRDs. In most LRDs, a high-column-density gas envelope with a covering factor near unity prevents broad Ly$\alpha$ photons from escaping. Broad Ly$\alpha$ can be observed in a minority of LRDs if photons escape directly through a lower-covering-factor envelope or through a small number of gaps in a clumpy and porous envelope.}
\label{fig:broad_lya_schematic}
\end{figure*}

Figure~\ref{fig:broad_lya_schematic} presents a schematic interpretation of LRD structure based on our results. The majority of LRDs, approximately 90\%, do not show broad Ly$\alpha$ emission and instead exhibit only narrow Ly$\alpha$ or no detected Ly$\alpha$ emission (Section~\ref{sec:broad_lya_fraction}). The width and strength of narrow Ly$\alpha$ are both consistent with those of actively star-forming galaxies such as EELGs (Sections~\ref{sec:lya_fwhm_distribution} and \ref{sec:lya_equivalent_width}), suggesting that the narrow emission is dominated by the host galaxy. One possible interpretation is that these LRDs are surrounded by a dense gas envelope with a covering factor close to unity. Broad Ly$\alpha$ photons propagating through such an envelope may undergo so much resonant scattering that they become too redshifted and broadened to be detected, or they may be destroyed through processes such as collisional transitions from the H~\textsc{i} $2p$ state to other states \citep{neufeld90, xji26, tang26}.

In contrast, the minority of LRDs with broad Ly$\alpha$ emission, approximately 10\%, appear to have an escape channel from the BLR \citep{xji26,tang26}. A covering factor below unity is one possible explanation. For example, if Abell2744-QSO1 has the covering factor of approximately 70\% estimated above, broad Ly$\alpha$ photons may escape directly through gaps in the envelope. Even if $f_{\mathrm{cov}}\sim1$, Ly$\alpha$ photons may escape from a clumpy and porous envelope by scattering from the surfaces of dense gas clumps. Escape without a substantial redshift requires the number of clumps along the line of sight to be sufficiently small \citep{neufeld90,gronke16,tang26}. Broad Ly$\alpha$ emission like that of Abell2744-QSO1 may therefore be visible when the column density is lower or when the characteristic gas-cloud size is larger, reducing the number of clumps along the line of sight. The small $f_{\mathrm{broad}}$ measured in this work suggests that LRDs satisfying these conditions, or having a low $f_{\mathrm{cov}}$, are uncommon. Instead, most LRDs appear to be nearly uniformly covered by a high-column-density envelope.

A remaining question is what distinguishes the few LRDs with broad Ly$\alpha$ emission, such as Abell2744-QSO1 and CANUCS-LRD-z8.6, from the rest of the population. One possibility is that they correspond to a particular evolutionary stage in which a strong outflow has partially disrupted the surrounding dense gas envelope. Fitting the Ly$\alpha$ profile of Abell2744-QSO1 from the SPURS observations \citep{tang26} with the zELDA thin-shell radiative-transfer models \citep{gurunglopez19, gurunglopez22} favors an expansion velocity of $V_{\mathrm{exp}}\sim400$--$600\ \mathrm{km\ s^{-1}}$, indicating a fast outflow. An outflow of this speed could blow out part of the envelope, lowering the covering factor to $f_{\mathrm{cov}}<1$ and creating the escape channels inferred above. Under this picture, broad Ly$\alpha$ emission would be a short-lived signature seen only while the outflow is actively clearing the envelope, which would naturally account for its rarity. We emphasize that this scenario remains speculative, because it rests on the two broad Ly$\alpha$ emitters currently known and on a simplified shell geometry that may not capture the structure of a clumpy envelope. Testing it will require theoretical work on how radiatively or mechanically driven outflows reshape the dense gas around LRD nuclei, together with larger samples of LRDs observed at $R\gtrsim1000$ to establish whether broad Ly$\alpha$ emission is systematically accompanied by outflow signatures.

\section{Summary}
\label{sec:summary}

In this work, we use VLT/MUSE integral-field spectroscopy and JWST/NIRSpec grating spectroscopy to present a demographic study of Ly$\alpha$ line profiles in LRDs at $z=3$--9. Our analysis and main results are summarized below:

\begin{itemize}
\item We analyze eight MUSE LRDs and 23 JWST/NIRSpec LRDs, corresponding to 27 distinct sources after accounting for overlap. Only Abell2744-QSO1 and CANUCS-LRD-z8.6 show broad Ly$\alpha$ emission. After accounting for the different spectral sensitivities, the representative broad Ly$\alpha$ emitter fraction $f_{\mathrm{broad}}$ at $L_{\mathrm{Ly}\alpha,\mathrm{broad}}\geq10^{42}\ \mathrm{erg\ s^{-1}}$ is $0.10_{-0.07}^{+0.12}$ for LRDs, larger than the 95\% upper limit of $0.02$ for SFGs.

\item We reproduce the relatively broad Ly$\alpha$ profile reported in the eight-source JWST stack of \citet{geris26}. In contrast, the larger 21-source JWST stack and the higher-resolution eight-source MUSE stack are best described by single-Gaussian profiles with $\mathrm{FWHM}_{\mathrm{Ly}\alpha}=619$ and $399\ \mathrm{km\ s^{-1}}$, respectively. We find no evidence for a hidden broad Ly$\alpha$ component associated with broad H$\alpha$ in either stack.

\item Apart from the two broad Ly$\alpha$ emitters, the Ly$\alpha$ FWHM distribution of LRDs is consistent with that of SFGs after accounting for the LSF uncertainty. The rest-frame Ly$\alpha$ equivalent width of the UV-normalized LRD stack is approximately four times higher than that of the SFG stack and is comparable to those of the JADES EELG stacks, suggesting that narrow Ly$\alpha$ is dominated by vigorous host-galaxy star formation.

\item We interpret the small $f_{\mathrm{broad}}$ and the high covering factor and column density inferred for Abell2744-QSO1 as evidence that most LRDs are nearly uniformly covered by a high-column-density gas envelope. Broad Ly$\alpha$ may escape only in the minority of LRDs with a lower-covering-factor or clumpy and porous envelope.
\end{itemize}

\begin{acknowledgments}
This work is based on observations taken by the MUSE-Wide Survey and MUSE Hubble Ultra Deep Field surveys as part of the MUSE Consortium. We thank the MUSE-Wide and MUSE Hubble Ultra Deep Field teams for making their reduced datacubes, one-dimensional spectra, and emission-line catalogs publicly available. This work is based on observations made with the NASA/ESA/CSA James Webb Space Telescope. The data were obtained from the Mikulski Archive for Space Telescopes at the Space Telescope Science Institute, which is operated by the Association of Universities for Research in Astronomy, Inc., under NASA contract NAS 5-03127 for JWST. These observations are associated with programs GTO 1180 (PI: Eisenstein), GTO 1181 (PI: Eisenstein), GTO 1208 (PI: Willott), GTO 1210 (PI: Luetzgendorf), GTO 1286 (PI: Luetzgendorf), GTO 1287 (PI: Luetzgendorf), GO 2561 (PIs: Labbe \& Bezanson), GO 3215 (PIs: Eisenstein \& Maiolino), GO 4287 (PIs: Mason \& Stark), GTO 4552 (PI: Stiavelli), GO 4750 (PI: Nakajima), GO 5943 (PIs: Papovich, Hutchison, \& Hu), GO 8018 (PI: Lin), GO 8060 (PIs: Egami, Maiolino, \& Rest), and GO 9214 (PIs: Mason \& Stark). The authors acknowledge the UNCOVER, JOF, DIVER, GO 8060, and SPURS teams, led by PIs Labbe, Bezanson, Eisenstein, Maiolino, Lin, Egami, Rest, Mason, \& Stark, for developing their observing programs with a zero-exclusive-access period. Some of the data products presented herein were retrieved from the Dawn JWST Archive (DJA). DJA is an initiative of the Cosmic Dawn Center (DAWN), which is funded by the Danish National Research Foundation under grant DNRF140. We thank DAWN for providing the reduced NIRSpec data. We thank the JADES team for making its spectroscopic catalog and reduced spectra publicly available, and the JADES, UNCOVER, and CANUCS teams for making their photometric catalogs publicly available. YK acknowledges support by KAKENHI (26KJ0960) through Japan Society for the Promotion of Science (JSPS), JSR Fellowship, and FoPM, WINGS Program, the University of Tokyo. MO acknowledges support from the World Premier International Research Center Initiative (WPI Initiative), MEXT, Japan, the joint research program of the Institute for Cosmic Ray Research (ICRR), the University of Tokyo, and KAKENHI (21H04467, 25H00674) through JSPS. HY acknowledges support by KAKENHI (25KJ0832) through JSPS. YH acknowledges support from the JSPS Grant-in-Aid for Scientific Research (24H00245) and the JSPS International Leading Research (22K21349). TK acknowledges support by KAKENHI (26KJ1232) through JSPS. MN is supported by JSPS KAKENHI Grant No. 25KJ0828. The authors acknowledge the use of Codex (OpenAI, GPT 5.5 and 5.6) and Claude Code (Anthropic, Opus 5 and Sonnet 5) to assist with language editing and code development. All AI-assisted outputs were carefully reviewed and validated by the authors. The authors take full responsibility for all analyses, interpretations, and conclusions presented in this work.
\end{acknowledgments}

\facilities{VLT, JWST}
\software{Astropy \citep{astropy13, astropy18, astropy22}, emcee \citep{foremanmackey13}, GetDist \citep{lewis25getdist}, grizli \citep{brammer21grizli}, Matplotlib \citep{hunter07}, MPDAF \citep{bacon16mpdaf}, msaexp \citep{brammer22msaexp}, msafit \citep{degraaff24msafit}, NumPy \citep{harris20}, photutils \citep{bradley25photutils}, PyNeb \citep{luridiana15}, SciPy \citep{virtanen20}, SpectRes \citep{carnall17}, zELDA \citep{gurunglopez19, gurunglopez22}}

\appendix
\section{Type 1 and Type 2 AGN Samples}
\label{sec:agn_sample_tables}

The spectroscopic Type 1 and Type 2 AGN comparison samples defined in Section~\ref{sec:non_lrd_agn} are listed in Tables~\ref{tab:type1_agn_sample} and \ref{tab:type2_agn_sample}, respectively.

\begin{deluxetable*}{llcccccccc}[h]
\tabletypesize{\scriptsize}
\tablecaption{Type 1 AGN Spectroscopic Sample\label{tab:type1_agn_sample}}
\tablehead{
\colhead{Field} & \colhead{ID} & \colhead{PID} & \colhead{R.A.} & \colhead{Decl.} & \colhead{$z_{\mathrm{sys}}$} & \colhead{$M_{\mathrm{UV}}$} & \colhead{$F_{\mathrm{Ly}\alpha}$} & \colhead{$\Delta v_{\mathrm{Ly}\alpha}$} & \colhead{$\mathrm{FWHM}_{\mathrm{Ly}\alpha}$} \\
\colhead{(1)} & \colhead{(2)} & \colhead{(3)} & \colhead{(4)} & \colhead{(5)} & \colhead{(6)} & \colhead{(7)} & \colhead{(8)} & \colhead{(9)} & \colhead{(10)} \\
\colhead{} & \colhead{} & \colhead{} & \colhead{(deg)} & \colhead{(deg)} & \colhead{} & \colhead{} & \colhead{($10^{-18}\,\mathrm{erg\,s^{-1}\,cm^{-2}}$)} & \colhead{($\mathrm{km\,s^{-1}}$)} & \colhead{($\mathrm{km\,s^{-1}}$)}
}
\startdata
GOODS-S & GS-30148179 & 1286 & 53.142075 & -27.779852 & 5.9212 & -20.2 & $<2.89$ &  &  \\
GOODS-S & GS-20030333 & 1286 & 53.053722 & -27.877890 & 7.8906 & -19.5 & $1.64^{+0.45}_{-0.52}$ & $<348$ & $706^{+200}_{-300}$ \\
GOODS-S & GS-164055 & 1286 & 53.081680 & -27.888569 & 7.3955 & -18.7 & $<2.06$ &  &  \\
GOODS-N & GN-77652 & 1181 & 189.293230 & 62.199013 & 5.2288 & -18.4 & $<4.42$ &  &  \\
GOODS-N & GN-62309 & 1181 & 189.248960 & 62.218360 & 5.1725 & -18.8 & $<4.68$ &  &  \\
GOODS-N & GN-38509 & 1181 & 189.091480 & 62.228115 & 6.6757 & -19.0 & $3.26^{+0.78}_{-0.88}$ & $373^{+200}_{-200}$ & $863\pm300$ \\
GOODS-N & GN-954 & 1181 & 189.151950 & 62.259644 & 6.7607 & -19.5 & $5.30^{+0.78}_{-0.88}$ & $<196$ & $620^{+100}_{-200}$ \\
\enddata
\tablecomments{(1) Field. (2) Source ID. (3) JWST/NIRSpec program ID. (4), (5) R.A. and Decl. (6) Systemic redshift. (7) Rest-frame UV absolute magnitude. (8) Ly$\alpha$ flux. (9) Ly$\alpha$ velocity offset. (10) Ly$\alpha$ FWHM. Upper limits on $F_{\mathrm{Ly}\alpha}$ and $\Delta v_{\mathrm{Ly}\alpha}$ are the 95\% and 68\% upper limits, respectively.}
\end{deluxetable*}

\begin{deluxetable*}{llcccccccc}[h]
\tabletypesize{\scriptsize}
\tablecaption{Type 2 AGN Spectroscopic Sample\label{tab:type2_agn_sample}}
\tablehead{
\colhead{Field} & \colhead{ID} & \colhead{PID} & \colhead{R.A.} & \colhead{Decl.} & \colhead{$z_{\mathrm{sys}}$} & \colhead{$M_{\mathrm{UV}}$} & \colhead{$F_{\mathrm{Ly}\alpha}$} & \colhead{$\Delta v_{\mathrm{Ly}\alpha}$} & \colhead{$\mathrm{FWHM}_{\mathrm{Ly}\alpha}$} \\
\colhead{(1)} & \colhead{(2)} & \colhead{(3)} & \colhead{(4)} & \colhead{(5)} & \colhead{(6)} & \colhead{(7)} & \colhead{(8)} & \colhead{(9)} & \colhead{(10)} \\
\colhead{} & \colhead{} & \colhead{} & \colhead{(deg)} & \colhead{(deg)} & \colhead{} & \colhead{} & \colhead{($10^{-18}\,\mathrm{erg\,s^{-1}\,cm^{-2}}$)} & \colhead{($\mathrm{km\,s^{-1}}$)} & \colhead{($\mathrm{km\,s^{-1}}$)}
}
\startdata
GOODS-S & GS-13176 & 1210 & 53.121754 & -27.797638 & 5.9357 & -19.8 & $8.89\pm0.47$ & $<73.7$ & $338\pm22$ \\
GOODS-S & GS-13577 & 1210 & 53.130035 & -27.778390 & 5.5667 & -19.4 & $<5.56$ &  &  \\
GOODS-S & GS-5173 & 1210 & 53.156826 & -27.767162 & 7.9813 & -18.8 & $1.45\pm0.21$ & $176^{+100}_{-50}$ & $423^{+80}_{-70}$ \\
GOODS-S & GS-15338 & 1210 & 53.115320 & -27.772892 & 5.0774 & -19.3 & $<15.9$ &  &  \\
GOODS-S & GS-56849 & 1210 & 53.113530 & -27.772816 & 5.8154 & -18.1 & $6.25\pm0.97$ & $149^{+100}_{-100}$ & $424\pm90$ \\
GOODS-S & GS-99671 & 3215 & 53.126640 & -27.817732 & 5.9222 & -18.0 & $<1.23$ &  &  \\
GOODS-S & GS-201127 & 3215 & 53.166843 & -27.804132 & 5.8295 & -18.6 & $<5.22$ &  &  \\
GOODS-S & GS-202208 & 3215 & 53.164074 & -27.799720 & 5.4470 & -19.9 & $<3.03$ &  &  \\
GOODS-S & GS-208643 & 3215 & 53.130190 & -27.778366 & 5.5642 & -18.0 & $<8.85$ &  &  \\
\enddata
\tablecomments{(1) Field. (2) Source ID. (3) JWST/NIRSpec program ID. (4), (5) R.A. and Decl. (6) Systemic redshift. (7) Rest-frame UV absolute magnitude. (8) Ly$\alpha$ flux. (9) Ly$\alpha$ velocity offset. (10) Ly$\alpha$ FWHM. Upper limits on $F_{\mathrm{Ly}\alpha}$ and $\Delta v_{\mathrm{Ly}\alpha}$ are the 95\% and 68\% upper limits, respectively.}
\end{deluxetable*}

\bibliography{Kageura26b}{}
\bibliographystyle{aasjournalv7}



\end{document}